 \documentclass[12pt]{article}
\usepackage{amsmath}
\usepackage{amssymb}
 \usepackage{epsfig}

\newcounter{multieqs}

\newcommand{\be}{\begin{equation}}
\newcommand{\ee}{\end{equation}}
\newcommand{\eq}[1]{(\ref{#1})}
\newcommand{\bit}{\begin{itemize}}  \newcommand{\eit}{\end{itemize}}
\newcommand{\ben}{\begin{enumerate}}  \newcommand{\een}{\end{enumerate}}

\newcommand{\bm}[1]{\mbox{\boldmath $#1$}}
\newcommand{\rf}[1]{(\ref{#1})}

\def\bd{\begin{document}}
\def\ed{\end{document}}
\def\bea{\begin{eqnarray}}
\def\eea{\end{eqnarray}}
\let\bm=\bibitem

\def\la{\langle}
\def\ra{\rangle}

\def\npb#1#2#3{Nucl. Phys. {\bf{B#1}} #3 (#2)}
\def\plb#1#2#3{Phys. Lett. {\bf{#1B}} #3 (#2)}
\def\prl#1#2#3{Phys. Rev. Lett. {\bf{#1}} #3 (#2)}
\def\prd#1#2#3{Phys. Rev. {D bf{#1}} #3 (#2)}
\def\cmp#1#2#3{Comm. Math. Phys. {\bf{#1}} #3 (#2)}
\def\cqg#1#2#3{Class. Quantum Grav. {\bf{#1}} #3 (#2)}
\def\nppsa#1#2#3{Nucl. Phys. B (Proc. Suppl.) {\bf{#1A}}#3 (#2)}
\def\ap#1#2#3{Ann. of Phys. {\bf{#1}} #3 (#2)}
\def\ijmp#1#2#3{Int. J. Mod. Phys. {\bf{A#1}} #3 (#2)}
\def\rmp#1#2#3{Rev. Mod. Phys. {\bf{#1}} #3 (#2)}
\def\mpla#1#2#3{Mod. Phys. Lett. {\bf A#1} #3 (#2)}
\def\jhep#1#2#3{J. High Energy Phys. {\bf #1} #3 (#2)}
\def\atmp#1#2#3{Adv. Theor. Math. Phys. {\bf #1} #3 (#2)}

\def\N{{\cal N}}
\def\sst{\scriptscriptstyle}
\def\thetabar{\bar\theta}
\def\Tr{{\rm Tr}}
\def\one{\mbox{1 \kern-.59em {\rm l}}}

\def\a{\alpha}      \def\da{{\dot\alpha}}  \def\dA{{\dot A}}
\def\b{\beta}       \def\db{{\dot\beta}}
\def\g{\gamma}  \def\G{\Gamma}  \def\dc{{\dot\gamma}}
\def\d{\delta}  \def\D{\Delta}  \def\ddt{\dot\delta}
\def\e{\epsilon}
\def\ve{\varepsilon}
\def\uve{\upvarepsilon}
\def\f{\phi}    \def\F{\Phi}    \def\vvf{\f}
\def\vphi{\varphi}
\def\h{\eta}
\def\k{\kappa}
\def\l{\lambda} \def\L{\Lambda}
\def\m{\mu} \def\n{\nu}
\def\o{\omega}
\def\p{\pi} \def\P{\Pi}
\def\r{\rho}
\def\s{\sigma}  \def\S{\Sigma}
\def\t{\tau}
\def\th{\theta} \def\Th{\Theta} \def\vth{\vartheta}
\def\X{\Xeta}
\def\z{\zeta}

\def\na{\nabla}

\def\cA{{\mathscr A}} \def\cB{{\cal B}} \def\cC{{\cal C}}
\def\cD{{\cal D}} \def\cE{{\cal E}} \def\cF{{\cal F}}
\def\cG{{\cal G}} \def\cH{{\cal H}} \def\cI{{\cal I}}
\def\cJ{{\mathscr J}} \def\cK{{\cal K}} \def\cL{{\cal L}}
\def\cM{{\cal M}} \def\cN{{\cal N}} \def\cO{{\cal O}}
\def\cP{{\cal P}} \def\cQ{{\cal Q}} \def\cR{{\cal R}}
\def\cS{{\cal S}} \def\cT{{\cal T}} \def\cU{{\cal U}}
\def\cV{{\cal V}} \def\cW{{\cal W}} \def\cX{{\cal X}}
\def\cY{{\cal Y}} \def\cZ{{\cal Z}}

\def\ua{\underline{\alpha}}
\def\uc{\underline{\phantom{\alpha}}\!\!\!\gamma}
\def\um{\underline{\mu}}
\def\ud{\underline\delta}
\def\ue{\underline\epsilon}
\def\una{\underline a}\def\unA{\underline A}
\def\unb{\underline b}\def\unB{\underline B}
\def\unc{\underline c}\def\unC{\underline C}
\def\und{\underline d}\def\unD{\underline D}
\def\une{\underline e}\def\unE{\underline E}
\def\unf{\underline{\phantom{e}}\!\!\!\! f}\def\unF{\underline F}
\def\unm{\underline m}\def\unM{{\underline M}}
\def\unn{\underline n}\def\unN{{\underline N}}
\def\unp{\underline{\phantom{a}}\!\!\! p}\def\unP{\underline P}
\def\unq{\underline{\phantom{a}}\!\!\! q}
\def\unQ{\underline{\phantom{A}}\!\!\!\! Q}
\def\unH{\underline{H}}

\def\As {{A \hspace{-6.4pt} \slash}\;}
\def\bs {{b \hspace{-6.4pt} \slash}\;}
\def\Ds {{D \hspace{-6.4pt} \slash}\;}
\def\Gts {{\Gt \hspace{-6.4pt} \slash}\;}
\def\ds {{\del \hspace{-6.4pt} \slash}\;}
\def\ss {{\s \hspace{-6.4pt} \slash}\;}
\def\ks {{ k \hspace{-6.4pt} \slash}\;}
\def\ps {{p \hspace{-6.4pt} \slash}\;}
\def\xs {{x \hspace{-6.4pt} \slash}\;}
\def\pas {{{p_1} \hspace{-6.4pt} \slash}\;}
\def\pbs {{{p_2} \hspace{-6.4pt} \slash}\;}
\def\cFs {{{\cal F} \hspace{-6.4pt} \slash}\;}
\def\Dss {{D \hspace{-7.5pt} \slash}\;}
\def\dss {{\del \hspace{-7.0pt} \slash}\;}

\def\Ah{{\hat{A}}}
\def\Dh{{\hat{D}}}
\def\Gh{{\hat{G}}}
\def\Fh{{\hat{F}}}
\def\Ih{{\hat{I}}}
\def\Jh{{\hat{J}}}
\def\Kh{{\hat{K}}}
\def\Lh{{\hat{L}}}
\def\Ph{{\hat{P}}}
\def\Rh{{\hat{R}}}
\def\Vh{{\hat{V}}}
\def\Xh{{\hat{X}}}

\def\ah{{\hat{\a}}}
\def\bh{{\hat{\b}}}
\def\gh{{\hat{\g}}}
\def\dh{{\hat{\d}}}
\def\rh{{\hat{\r}}}
\def\hh{\hat{h}}
\def\uh{\hat{u}}
\def\xh{\hat{x}}
\def\yh{\hat{y}}
\def\ph{\hat{p}}
\def\xih{\hat{\xi}}
\def\chih{\hat{\chi}}
\def\Psih{\hat{\Psi}}
\def\phih{\hat{\phi}}

\def\psit{\tilde{\psi}}
\def\Psit{\tilde{\Psi}}
\def\Psibt{\tilde{\bar{Psi}}}

\def\st{\tilde{\sigma}}

\def\delt{\tilde{\delta}}
\def\Phit{\tilde{\Phi}}
\def\Phitb{\overline{\tilde{Phi}}}
\def\tht{\tilde{\th}}
\def\lt{\tilde{\l}}
\def\chit{\tilde{\chi}}
\def\phit{\tilde{\phi}}

\def\At{\tilde{A}}
\def\Bt{\tilde{B}}
\def\Ct{\tilde{C}}
\def\Dt{\tilde{D}}
\def\Et{\tilde{E}}
\def\Ft{\tilde{F}}
\def\Gt{\tilde{G}}
\def\Ht{\tilde{H}}
\def\It{\tilde{I}}
\def\Jt{\tilde{J}}
\def\Qt{\tilde{Q}}
\def\Rt{\tilde{R}}
\def\Mt{\tilde{M }}
\def\Nt{\tilde{N}}
\def\St{\tilde{S}}
\def\Vt{\tilde{V}}
\def\Xt{\tilde{X}}
\def\at{\tilde{a}}
\def\ct{\tilde{c}}
\def\dt{\tilde{d}}
\def\htt{\tilde{h}}
\def\ft{\tilde{f}}
\def\gt{\tilde{g}}
\def\pt{\tilde{p}}
\def\qt{\tilde{q}}
\def\vt{\tilde{v}}
\def\nt{\tilde{n}}
\def\ut{\tilde{u}}
\def\wt{\tilde{w}}
\def\zt{\tilde{z}}
\def\xt{\tilde{x}}
\def\yt{\tilde{y}}
\def\Psit{\tilde{\Psi}}
\def\vphit{\tilde{\varphi}}
\def\tD{\tilde{\D}}

\def\eb{\bar{\epsilon}}
\def\delb{\bar{\partial}}
\def\thb{\bar{\theta}}
\def\mub{\bar{\mu}}
\def\lamb{\bar{\l}}
\def\psib{\bar{\psi}}
\def\sb{\bar{\sigma}}
\def\xib{\bar{\xi}}
\def\chib{\bar{\chi}}

\def\Psib{\bar{\Psi}}
\def\Phib{\bar{\Phi}}
\def\Lamb{\bar{\Lambda}}
\def\Sb{{\overline \Sigma}}
\def\cb{\bar{c}}
\def\hb{\bar{h}}
\def\qb{\bar{q}}
\def\wb{\bar{w}}
\def\ub{\bar{u}}
\def\zb{{\bar{z}}}
\def\Hb{\bar{H}}
\def\Qb{{\bar Q}}
\def\Omegab{\overline{\Omega}}
\def\ob{\overline{\omega}}

\def\Ab{{\overline A}} \def\Bb{{\overline B}} \def\Cb{{\overline C}}
\def\Db{{\overline D}} \def\Eb{{\overline E}} \def\Fb{{\overline F}}
\def\Gb{{\overline G}}
\def\Ib{{\overline I}}
\def\Jb{{\overline J}} \def\Kb{{\overline K}} \def\Lb{{\overline L}}
\def\Mb{{\overline M}} \def\Nb{{\overline N}} \def\Ob{{\overline O}}
\def\Pb{{\overline P}}  \def\Rb{{\overline R}}
 \def\Tb{{\overline T}} \def\Ub{{\overline U}}
\def\Vb{{\overline V}} \def\Wb{{\overline W}} \def\Xb{{\overline X}}
\def\Yb{{\overline Y}} \def\Zb{{\overline Z}}

\def\fb{{\overline f}}
\def\gb{{\overline g}}
\def\mb{{\overline m}}
\def\lb{{\overline l}}
\def\yb{{\overline y}}

\def\ldel{{\overleftarrow{\del}}}
\def\rdel{{\overrightarrow{\del}}}
\def\ldeldel{{\overleftarrow{\del^2}}}
\def\rdeldel{{\overrightarrow{\del^2}}}
\def\ldelb{{\overleftarrow{\bar{\del}}}}
\def\rdelb{{\overrightarrow{\bar{\del}}}}

\def\ba{{\bf a}}
\def\bk{{\bf k}}
\def\bl{{\bf l}}
\def\bp{{\bf p}}
\def\bq{{\bf q}}
\def\br{{\bf r}}
\def\bt{{\bf t}}
\def\bu{{\bf u}}
\def\bv{{\bf v}}
\def\bx{{\bf x}}
\def\by{{\bf y}}
\def\bA{{\bf A}}
\def\bB{{\bf B}}
\def\bR{{\bf R}}
\def\bV{{\bf V}}

\def\bz{{\boldsymbol{\zeta}}}

\def\bone{{\bf 1}}

\def\va{{\vec a}}
\def\vk{{\vec k}}
\def\vp{{\vec p}}
\def\vq{{\vec q}}
\def\vx{{\vec x}}
\def\vy{{\vec y}}
\def\vu{{\vec u}}
\def\vv{{\vec v}}
\def \vH{{\vec H}}
\def \vg{{\vec g}}

\def\vs{{\vec \sigma}}
\def\vtau{{\vec \tau}}

\newcommand{\ov}[1]{\overrightarrow{#1}}

\def\frA{\mathfrak{A}}
\def\frB{\mathfrak{B}}
\def\frC{\mathfrak{C}}
\def\frD{\mathfrak{D}}
\def\frE{\mathfrak{E}}
\def\frF{\mathfrak{F}}
\def\frG{\mathfrak{G}}
\def\frH{\mathfrak{H}}
\def\frM{\mathfrak{M}}
\def\frN{\mathfrak{N}}
\def\frR{\mathfrak{R}}
\def\frW{\mathfrak{W}}

\def\fra{\mathfrak{a}}
\def\frb{\mathfrak{b}}
\def\frf{\mathfrak{f}}
\def\frg{\mathfrak{g}}
\def\frh{\mathfrak{h}}
\def\frl{\mathfrak{l}}
\def\frs{\mathfrak{s}}
\def\fri{\mathfrak{i}}
\def\frj{\mathfrak{j}}

\def\ma{\mathfrak{a}}
\def\mg{\mathfrak{g}}
\def\mh{\mathfrak{h}}
\def\mR{\mathfrak{R}}
\def\mN{\mathfrak{N}}

\newcommand{\nn}{{\nonumber}}

\def\d{\delta}\def\D{\Delta}\def\ddt{\dot\delta}

\def\pa{\partial} \def\del{\partial}
\def\xx{\times}
\def\uno{\mbox{1 \kern-.59em {\rm l}}}

\def\trp{^{\top}}
\def\inv{^{-1}}
\def\dag{\dagger}
\def\pr{^{\prime}}

\def\rar{\rightarrow}
\def\lar{\leftarrow}
\def\lrar{\leftrightarrow}

\newcommand{\0}{\,\!}      
\def\one{1\!\!1\,\,}
\def\im{\imath}
\def\jm{\jmath}

\newcommand{\tr}{\mbox{tr}}
\newcommand{\slsh}[1]{/ \!\!\!\! #1}

\def\vac{|0\rangle}
\def\lvac{\langle 0|}

\def\hlf{\frac{1}{2}}
\def\ove#1{\frac{1}{#1}}
\newcommand{\hot}[1]{\frac{#1}{2}}

\def\Box{\square}
\def\CC {\mathbb{C}}
\def\FF {\mathbb{F}}
\def\RR{\mathbb{R}}
\def\NN{\mathbb{N}}
\def\ZZ{\mathbb{Z}}
\def\bb#1{{\bf #1}}
\def\bcomment#1{}
\def\bfhat#1{{\bf \hat{#1}}}
\def\VEV#1{\left\langle #1\right\rangle}

\newcommand{\ex}[1]{{\rm e}^{#1}} \def\ii{{\rm i}}

\newcommand{\lrbrk}[1]{\left(#1\right)}
\newcommand{\lrsbrk}[1]{\left[#1\right]}
\newcommand{\sfrac}[2]{{\textstyle\frac{#1}{#2}}}

\def\stw{{\sqrt{2}}}

\def\rf {{\rm f}}
\def\ri {{\rm i}}
\def\rj {{\rm j}}
\def\rn {{\rm n}}
\def\rk {{\rm k}}
\def\rl {{\rm l}}
\def\rr {{\rm r}}
\def\rs {{\scriptscriptstyle \rm S}}
\def\rt {{\scriptscriptstyle \rm T}}

\def\rQ {{\scriptscriptstyle \rm \cQ}}
\def\rR {{\scriptscriptstyle \rm \cR}}

\def\cQb{{\cal \Qb}}
\def\cRb{{\cal \Rb}}
\def\cWb{{\cal \Wb}}

\def\fd {{\rm N}}
\def\afd {{\overline{\rm N}}}

\def \II {I\hspace{-.1em}I\hspace{.1em}}
\def \IIA {\mbox{\II A\hspace{.2em}}}
\def \IIB {\mbox{\II B\hspace{.2em}}}
\def \gs {g^s}
\def \ls {\lambda^s}

\def \I {{\cal I}}
\def \qs {q\hspace{-.53em}/\hspace{.15em}}
\def \ks {k\hspace{-.53em}/\hspace{.15em}}
\def \YM {{\mbox{\tiny YM}}}
\def \gym {g_{\YM}}

\def \Lc {\L_c}
\def\IR{\relax{\rm I\kern-.18em R}}
\def \id {{\bf 1}}

\def\cci{\ell}
\def\ccj{\ell'}

\def\bbq{\pmb{q}}

\newcommand{\para}[1]{\vskip 0.1cm {\noindent{#1}} \vskip 0.1cm}
\newcommand{\parabf}[1]{\vskip 0.3cm {\noindent{\bf #1}} \vskip 0.0cm}
\newcommand{\parait}[1]{\vskip 0.3cm {\noindent{\it #1}} }
\newcommand{\paratt}[1]{\vskip 0.1cm {\noindent{\tt #1}} \vskip 0.1cm}
\newcommand{\parasl}[1]{\vskip 0.1cm {\noindent{\sl #1}} \vskip 0.1cm}
\newcommand{\parasf}[1]{\vskip 0.1cm {\noindent{\sf #1}} \vskip 0.1cm}
\newcommand{\parasc}[1]{\vskip 0.1cm {\noindent{\sc #1}} \vskip 0.1cm}
\newcommand{\paraun}[1]{\vskip 0.1cm {\noindent\underline{\sf #1}}}

\begin{document}

\begin{titlepage}
\begin{flushright}
\hfill{NCTS-TH/1903}
 \end{flushright}
\hfill 

\begin{center}

{\Large \bf
Memory Effect in Anti-de Sitter Spacetime}\\[10mm]


{\bf Chong-Sun Chu and Yoji Koyama}

{\itshape Physics Division, National Center for Theoretical
 Sciences,\\
 National Tsing-Hua University, Hsinchu, 30013, Taiwan\\
Department of Physics, National Tsing-Hua
University,  Hsinchu 30013, Taiwan}\\
{Email: cschu@phys.nthu.edu.tw, koyama811@gmail.com}
\end{center}

\begin{abstract}
  The geodesic deviation of a pair of test particles is an natural observable
  for the gravitational memory effect. Nevertheless
  in curved spacetime, this
  observable is plagued with various issues that needs to be clarified before
  one can extract the essential part that is related to the gravitational
  radiation.
  In this paper we consider the Anti deSitter space as an
  example and analyze this observable carefully.
  We show that by employing the Fermi Normal Coordinates around the geodesic
  of one of the particles
(i.e. the standard free falling frame attached to this particle),
   one can elegantly
  separate out the curvature contribution of the background spacetime
  to the geodesic deviation from the contribution of
the gravitational wave. The  
gravitational wave  memory obtained this way depends linearly and
locally on the retarded metric perturbation caused by the gravitational wave,
and, remarkably, it takes on exactly
the same formula \eq{memory-flat} as in the flat case. To determine the memory,
in addition to the standard tail contribution to the gravitational radiation,
one need to take into account of the contribution from the reflected
gravitational wave off the AdS boundary.
For general curved spacetime, our analysis suggests that the use of a certain
coordinate system adapted to the
local geodesic (e.g. the Fermi normal coordinates system in the AdS case) would
allow one to dissect the geodesic deviation of test particles and
extract the relevant
contribution to define the
memory due to gravitational radiation.
\end{abstract}

{\it Keywords: Memory Effect, Gravity, Anti-deSitter Space}

\end{titlepage}

\section{Introduction}

Passage of a  gravitational wave (GW)
can induce a permanent displacement in 
the relative separation of a pair of 
 test particles  which serves as a gravitational wave detector.
This phenomenon is known as 
the {\it gravitational  memory effect}
\cite{Zeldovich}.
In a flat background, the
net relative displacement $\Delta D_{\mu}$ between the test particles,
after passage of the gravitational radiation, is given by 
\be
\Delta D_{\mu}
=\frac12 \Delta h^{TT}_{\mu\nu}D^{\nu},
\label{memory-flat}
\ee 
where $D^{\nu}$ is the initial separation of 
the pair of test particles, and 
$
\Delta h^{TT}_{\mu\nu}
$
is the net change in the metric perturbations in 
the transverse-traceless gauge.
In flat spacetime, the memory formula \eq{memory-flat} is simple.
The gravitational memory $\D D_\m$ 
is  determined
entirely in terms of the metric perturbation, which, in linearized
gravity, can be solved in terms of the retarded Green function $G_R$.
Schematically, without going into details about  gauge fixing
and decomposition into irreducible components, the retarded Green function $G_R$
satisfies
\be
\Box_x G_R (x,x')= - \d^{(4)}(x,x'),
\ee
where $\Box$ is a second order differential operator in which the linearized
Einstein equation can be written as $\Box_x h_{\m\n} = -16 \pi G T_{\m\n}$.  
As a result, we have
\be
h_{\m\n}(x) = 16 \pi G \int d^4 x' G_R(x,x') T_{\m\n}(x')
\ee
and the study of the properties of the memory effect can be phrased entirely
in terms of the
properties of the retarded Green function.
For  gravitational
wave generated by a source in a localized region of spacetime,
$\Delta h^{TT}_{\mu\nu}$ 
is of Coulomb type \cite{Braginskii:1987}, i.e.
$\Delta h^{TT}_{\mu\nu} \sim 1/r$ at
large distance $r$ of the detector from the source. For example, the
collision or explosion of a collection of freely
moving particles produces
the perturbation
\cite{Braginskii:1987,Thorne:1992sdb}
\be
\Delta h^{TT}_{\mu\nu}
=\frac{1}{r}
\Delta
\left[
\sum^{N}_{A=1}
\frac{4M_{A}}{\sqrt{1-v_{A}^{2}}}
\left(
\frac{v_{A \mu}v_{A \nu}}
{1-v_{A}\cos\theta_{A}}
\right)^{TT}
\right]
.
\label{ret-flat}
\ee
where $M_{A}$ and $v_{A}$ are the mass and the velocity 
of each freely moving massive body, respectively, and 
$\theta_{A}$ is the angle between the source and the detector. 
$TT$ denotes the transverse-traceless part of the expression.
Memory effect
in flat space and its properties as
produced by various kinds of massive and massless 
particle sources  were recently discussed in
\cite{Bieri:2011zb,Bieri:2013gwa,Bieri:2013ada,Tolish:2014bka,Garfinkle:2017fre}.
Detectability of the memory effect 
from gravitational wave signals associated with
binary black hole mergers was discussed 
in \cite{Thorne:1992sdb,Wiseman:1991ss}.
Recent discussions about the detectability 
with LISA, pulsar timing arrays or LIGO
can be found in 
\cite{Favata:2010zu,Seto2009,Haasteren2010,Lasky:2016knh}.

We are interested in the memory effect in curved spacetime. There are various
motivations for this interest. First, our universe is curved and not flat. It was
deSitter like at the time of inflation. Currently it is described by a
FLRW spacetime. One can imagine that the observation of gravitational
memory effect may provide valuable information on the structure
of the universe at various stages of its
development. Theoretically, gravitational memory effect in flat space
is related to the asymptotic BMS symmetry \cite{Bondi:1962px,Sachs:1962wk}
and the infrared properties of gravity
\cite{Strominger:2014pwa,Strominger:2017zoo}.
It is interesting to understand how much of this story
may
carry through in a curved background spacetime.

While the memory effect in dS and FLRW spacetimes
has been extensively studied, memory effect in AdS space
has been considered much less in the literature 
(see, for example, \cite{Bieri:2015jwa,Chu:2016qxp,Hamada:2017gdg,
  Kehagias:2016zry,Tolish:2016ggo,Bieri:2017vni,
  Ashtekar:2015lxa,Ashtekar:2015ooa}; also
\cite{Mishra:2018axf} for 
a study of the memory effect in gauge theory).
Although the memory effect in AdS space
is currently less motivated from observational point of view
than that in dS and FLRW spacetimes,
theoretically it is an interesting subject.
First,
as we will demonstrate in this paper,
the study of memory effect in the
AdS spacetime
gives us useful insights
on better ways to think about and analysis the memory effect in general
curved spacetime. Besides,
the  existence of boundary in AdS also offer an interesting
opportunity to study how the reflected gravitational wave may affect the
gravitational memory. 
Moreover, the study of AdS memory
is potentially be related to 
the other gravitational phenomena such as 
asymptotic symmetries of spacetime
and AdS/CFT correspondence 
\cite{Maldacena:1997re}.
These are some of the reasons behind that form the main motivations of this work.

The study of gravitational memory in curved spacetime is however much more
complicated and a number of effects not occurring in the flat spacetime
needed to be
taken into account. {\bf 1.}
In curved spacetime, the simple result \eq{memory-flat} no longer
holds and the net change $\D D_\m$
in the geodesic separation of test particles has to be obtained from
solving the geodesic derivation equation.
Due to the presence of nontrivial spacetime curvature, $\D D_\mu$
generally involves integration over the history of the motion of the particles
and takes a much more complicated nonlocal form compared to \eq{memory-flat}.
In spacetime without boundary, one can gain huge simplification
by making the observation at large
distance and at large time so that a local un-integrated
expression for $\D D_\m$ is obtained
\footnote{See for example \cite{Chu:2016qxp} for the dS case
to see how the nonlocal expression
(equation (5) there) can be reduced to a local expression (equation (6)) there
at large time.}. This however does not work for the AdS space due to the
presence of boundary, and it seems that the expression of the
memory in AdS space will necessary
be much more complicated. 
{\bf 2.}  In a curved spacetime, the geodesic separation between
the pair of particles 
get contributions from both the  background curvature
(back scattering by the gravitational potential created by the background
curvature)
as well as the gravitational wave.
Therefore in order to have a proper
definition of memory due to gravitational wave, it is important to
separate the contribution of the background curvature from the contribution
of the gravitational wave.
In  this work, we show that the above two issues can be resolved elegantly
by adopting a particular choice of coordinate system
to make the observation of memory. 
{\bf 3.}
Another subtlety in curved spacetime is the presence of the {\it tail term} 
in the retarded Green function, i.e. propagation of
gravitational waves less than the speed of light, besides the
propagation of gravitational waves at the speed of light which we will
call the {\it direct term}.
Let us comment on these contributions to the memory in
various spacetime 
such as the dS spacetime \cite{Bieri:2015jwa,Hamada:2017gdg},
decelerating FLRW spacetime with future null infinity \cite{Kehagias:2016zry}, 
spatially flat FLRW cosmologies \cite{Tolish:2016ggo}, and the
$\Lambda$CDM cosmologies \cite{Bieri:2017vni} that has been considered in the
literature. 
For the spatially flat FLRW spacetime \cite{Tolish:2016ggo} with localized
source,
 the authors  proposed 
 to continue to use the flat space criteria of the presence of a derivative of a
 delta function in the Riemann curvature as a way to 
characterize the gravitational memory for the spatially flat FLRW spacetime.
By construction, the tail contribution was excluded in this characterization
of the memory.
For the decelerating FLRW spacetime \cite{Kehagias:2016zry},
it was found that the tail term in
 the retarded potential is subleading in the $1/r$ expansion at the
 future null infinity and the memory effect is given entirely by the
 direct term \cite{Kehagias:2016zry}. In these two cases, 
 the direct term contribution to  
the memory effect is simple and can be written in the same form 
as the flat case
up to an multiplicative factor of
an inverse power of the scale factor $a(\tau)$ 
at the time of detection
 when the source and the detector are placed at the
 same proper distance as that in flat space 
 at the time of gravitational wave emission. 
 On the other hand, for the deSitter space,
it was  found that for even dimensions higher than 2
\cite{Ashtekar:2015lxa,Ashtekar:2015ooa,Chu:2016qxp},
the gravitational wave tail contributes significantly to the memory effect
\footnote{It is known that
the tail term vanishes for odd dimensional deSitter spacetime.}.
Note that in this case the retarded potential is expanded in terms of 
the conformal time $\eta$, instead of $1/r$, 
in order to approach the future infinity \cite{Ashtekar:2015ooa}. 
At $\eta=0$, the tail and direct contributions 
are equal in size but opposite in sign 
and thus they cancel out each other, leaving the higher order terms in $H$,
in the future infinity. From these results, one can learn the lesson that
both the direct term and the tail term depend rather
sensitively on the
asymptotic geometry of the spacetime where the observation of the memory is made.
In this regards, since AdS  spacetime has a boundary and
it's asymptotics is completely different.
The determination of both the direct and the tail contributions
at generic finite location in AdS space is another motivation of this work.
 {\bf 4.} AdS space is also special since it has a boundary and
 reflection may occur, and the reflected wave may affect the observed
 memory effect in significant way.
We will show in this paper that, depending on the location of
the detector,
the reflected the gravitational wave
may make a significant contribution to the observed
gravitational memory.

In this work, we study memory effect 
on 4-dimensional AdS space in the Poincar\'e coordinates.
We restrict ourselves to the linear order 
of metric perturbation around the vacuum AdS space. 
As the background spacetime is nontrivial, the geodesic separation $D^\m$
is a gauge dependent quantity and it is necessary to specify an observer
which made the result of memory as transparent physically as possible.
We find that  by employing the Fermi Normal Coordinates around the
geodesic of one of the particles,
one can elegantly
  separate out the curvature contribution of the background spacetime
  from those of the gravitational wave. Remarkably, the obtained 
  gravitational wave memory takes on exactly
  the same formula \eq{memory-flat} as in the flat case and has a simple
  local and factorized dependence on the retarded metric perturbation
  caused by the gravitational wave. This simple formula 
  allows us to
  determine the memory entirely in terms of the waveform of the
  retarded gravitational radiation.
As we mentioned above, in AdS space 
the retarded propagator of metric perturbation
contains a tail term.
Aside from the tail term,
AdS space has a distinctive feature 
that the gravitational waves reach
the infinity (AdS boundary) in a finite time
and then get reflected back to the bulk spacetime.
As a consequence, in addition to the original gravitational wave, 
there will also be a gravitational wave reflected at the AdS boundary, each
accompanied by its respective tail term. The
net
memory is given by the sum of all these contributions.

  For general curved spacetime, our analysis suggests that the use of a certain
coordinate system that is adapted to the
local geodesic (e.g. the Fermi normal coordinates system in the AdS case) would
allow one to dissect the geodesic deviation of test particles and
extract the relevant
contribution to define the memory effect
due to gravitational radiation. This is an interesting
direction to further explore \cite{work}.

The organization of this article is as follows.
As advertised, the
retarded metric perturbation plays an important role in determining the memory in
AdS space, therefore we will first start in section 2 with 
the construction of the retarded solution for the
linear metric perturbation in AdS spacetime. Following Wald and Tolish
\cite{Tolish:2014bka},
we will consider 
localized energy source as an example and work out the retarded wave solution.
In section 3.1, we consider the use of the Fermi normal coordinates and find that
the background curvature contribution to the geodesic deviation can be easily
disentangled and subtracted away. The remaining part of the geodesic deviation
\eq{gdesolution} depends on the retarded gravitational wave
linearly and locally, and in fact takes on exactly  the same form \eq{memory-flat}
as in the flat case.
And this is true for any finite time.
This is quite remarkable and
is one of the main results of this work.
In section 3.2, we take into account of the tail term and the reflected wave and
analyze the memory effect.  
As the reflected wave plays an important role, the
effect of memory depends crucially on whether the detector is receiving the
reflected wave or not. We show that for an observer that receive the reflected
wave, the memory effect got canceled out completely. Hence it is interesting that
memory effect is different for different locations of the
gravitational wave detector in the AdS spacetime.  
In section 3.3, we consider the asymptotic form of the memory near the
AdS boundary and find that it has a delta function singularity
localized on the lightcone from the source. 
In section 3.4 and 3.5, we construct the AdS shock wave by taking a certain
limit of our perturbed metric. We find that the velocity-memory for AdS shock wave
picks up a kink contribution  $u \th(u)$ in addition to
the jump $\th(u)$ and  pulse term $ \d(u)$ which are present in the flat case.
Section 4 contains our conclusion and some discussions. Some of the
more technical details of the analysis are contained in the appendices.

\section{Linear Perturbation in AdS Space}
\label{sectionperturbation}

Consider a $n$-dimensional AdS space. In the Poincar\'e coordinates,
the line element is given by
\bea
ds^{2}
=
{\bar g}_{\mu\nu}dx^{\mu}dx^{\nu}
=
\frac{L^{2}}{y^{2}}(-dt^{2}+(dx^{i})^{2}+dy^{2}),
\quad {\bar g}_{{\mu\nu}}=\frac{L^{2}}{y^{2}}\eta_{\mu\nu},
\label{ads}
\eea
with $-\infty<t,x^{i}<\infty$, $0<y<\infty$.
Here the subscripts $i,j=1,\dots,n-2$ 
and $\mu,\nu=0,\dots,n-2,y$,
and $L$ is the AdS radius. 
The AdS boundary is located at $y=0$ 
and the AdS horizon is at $y=\infty$.
Below we will use the notation for indices
such that $a,b=0,\dots,n-2$
and $r,s=1,\dots,n-2,y$.

\subsection{Retarded solution}

We consider perturbation  $\gamma_{\mu\nu}$ around 
the background AdS metric,
\be
g_{\mu\nu}={\bar g}_{\mu\nu}+\gamma_{\mu\nu}.
\ee
It is convenient to introduce the perturbation $\psi_{\mu\nu}$ 
from $\gamma_{\mu\nu}$ as defined by
\be
\psi_{\mu\nu}
= \gamma_{\mu\nu}-\frac12 {\bar g}_{\mu\nu}\gamma, \quad 
\gamma :={\bar g}^{\mu\nu}\gamma_{\mu\nu}.
\ee
We impose the following gauge conditions analogous to 
the one adopted in de-Sitter space \cite{deVega:1998ia}
\be
{\bar \nabla}_{\nu}\psi^{\nu}_{\ \mu}=-\frac{2}{y}\psi^{y}_{\ \mu},
\label{gauge}
\ee
where ${\bar \nabla}_{\mu}$ is the covariant derivative 
with respect to ${\bar g}_{\mu\nu}$. 
In terms of $\psi^{\nu}_{\ \mu}$, 
the linearized Einstein equation with a source is
\be
\partial^{2}\psi^{\nu}_{\ \mu}
-\frac{n-2}{y}\partial_{y}\psi^{\nu}_{\ \mu}
+\frac{1}{y^{2}}\Big(
(n-2)
(\delta^{\nu}_{\ y}\psi^{y}_{\ \mu}
+\delta^{y}_{\ \mu}\psi^{\nu}_{\ y})
-2\delta^{\nu}_{\ y}\delta^{y}_{\ \mu}\psi
\Big)
=
-\frac{16\pi G L^{2}}{y^{2}}T^{\nu}_{\ \mu},
\label{leinpsi}
\ee
where $\psi:={\bar g}^{\mu\nu}\psi_{\mu\nu}$,
$\partial^{2}
=\eta^{\rho\sigma}\partial_{\rho}\partial_{\sigma}$, 
$G$ is the Newton constant and 
$T^{\nu}_{\ \mu}$ stands for 
perturbative matter energy-momentum tensor.

It is convenient to introduce a re-scaled perturbation 
where its indices are raised and lowered by
$\eta_{\mu\nu}$,
\be
\chi_{\mu\nu}
:=\frac{y^{2}}{L^{2}}{\bar g}_{\mu\rho}\psi^{\rho}_{\ \nu}
=\eta_{\mu\rho}\psi^{\rho}_{\ \nu},
\quad
\chi :=\eta^{\mu\nu}\chi_{\mu\nu},
\ee
and in which the original perturbation 
$\gamma_{\mu\nu}$ can be written as 
\be
\gamma_{\mu\nu}
=\frac{L^{2}}{y^{2}}
\left(\chi_{\mu\nu}
-\frac12 \eta_{\mu\nu}\chi^{\rho}_{\ \rho}\right).
\ee
$\chi_{\m\n}$ satisfies the linearized Einstein equation
\be
\partial^{2}\chi_{\mu\nu}
-\frac{n-2}{y}\partial_{y}\chi_{\mu\nu}
+\frac{1}{y^{2}}\Big(
(n-2)(\eta_{y\nu}\chi_{\mu y}+\eta_{\mu y}\chi_{y\nu})
-2\eta_{y\nu}\eta_{\mu y}\chi
\Big)
= -\frac{16 \pi G L^2}{y^2} T^\n{}_\m. 
\label{leinchi}
\ee
Defining 
${\tilde \chi}=\chi^{\rho}_{\ \rho}-(n-2)\chi_{yy}$,
the equation \eqref{leinchi} can be decomposed into 
three independent equations:
\bea
&&
\partial^{2}\chi_{ab}
-\frac{n-2}{y}\partial_{y}\chi_{ab}
= -\frac{16 \pi G L^2}{y^2} T_{ab},
\label{chilinear1}
\\
&&
\partial^{2}\chi_{ya}
-\frac{n-2}{y}\partial_{y}\chi_{ya}
+\frac{1}{y^{2}}
(n-2)\chi_{ya}
= -\frac{16 \pi G L^2}{y^2} T_{ya},
\label{chilinear2}
\\
&&
\partial^{2}{\tilde \chi}
-\frac{n-2}{y}\partial_{y}{\tilde \chi}
+\frac{2}{y^{2}}
(n-3){\tilde \chi}
=-\frac{16 \pi G L^2}{y^2} \tilde{T},
\label{chilinear3}
\eea
where $\tilde{T}: =T^\r{}_\r -(n-2)T_{yy}$ and the subscript $a,b=0,\dots,n-2$.
In this paper we are interested in the $n=4$ dimensional AdS spacetime. In this
case,
the retarded solutions to \eq{chilinear1}-\eq{chilinear3} are given by
\begin{align}
\chi_{ab}(x)
&=
16\pi G
\int d^{4}x'
\frac{L^{2}}{y'^{2}}
G^{\nu=3/2}_{R}(x,x')
T_{ab}(x'),
\label{retab}
\\
\chi_{ya}(x)
&=
16\pi G
\int d^{4}x'
\frac{L^{2}}{y'^{2}}
G^{\nu=1/2}_{R}(x,x')
T_{ya}(x'),
\label{retya}
\\
{\tilde \chi}(x)
&=
16\pi G
\int d^{4}x'
\frac{L^{2}}{y'^{2}}
G^{\nu= - 1/2}_{R}(x,x')
{\tilde T}(x'), \quad
{\tilde T}=\eta^{\m\n}T_{\m\n}- 2 T_{yy}.
\label{rettilde}
\end{align}
where the retarded propagator $G^\n_R$ satisfy  the differential equations
\bea
&&
\left(\partial^{2}
-\frac{2}{y}\partial_{y}
\right)
G^{\nu=3/2}_{R}(x,x')
=
-\frac{y^{2}}{L^{2}}\delta^{(4)}(x-x'),
\\
&&
\left(\partial^{2}
-\frac{2}{y}\partial_{y}
+\frac{2}{y^{2}}
\right)
G^{ \nu=\pm 1/2}_{R}(x,x')
=
-\frac{y^{2}}{L^{2}}\delta^{(4)}(x-x').
\eea
Here the retarded propagator in AdS${}_4$ with index $\n$ is given by 
 \cite{Danielsson:1998wt} 
\begin{align}
G^{\nu}_{R}(w,w')
&=
-\frac{\theta(t-t'-|{\bf x}-{\bf x}'|)}{4\pi L^{2}}
\Bigg[
\frac{d}{dz_{+}}
\Big(
(\theta(1-z_{+})
-\theta(-1-z_{+})
)
P_{\nu-1/2}(z_{+})
\Big)
\notag\\
&\hspace{4cm}
+
2\cos(\nu\pi)
\frac{d}{dz_{-}}
\Big(\theta(z_{-}-1)
Q_{\nu-1/2}(z_{-})\Big)
\Bigg]
,
\label{gret4d}
\end{align}
where 
$P_{\nu-1/2}(z_{+})$ and $Q_{\nu-1/2}(z_{-})$ 
are the Legendre functions 
of the first and second kind respectively, 
and $z_{\pm}$ are given by
\be
z_{\pm}
=
\pm\frac{-(t-t')^{2}+|{\bf x}-{\bf x}'|^{2}+y^{2}+y'^{2}}{2yy'},
\qquad
|{\bf x}-{\bf x}'|=\sqrt{(x-x')_{i}(x-x')^{i}}.
\ee
For $\nu=3/2$ and 
$\nu = \pm 1/2$, 
$\cos(\nu\pi)=0$
 and we get the simple expressions
\begin{align}
G^{\nu=3/2}_{R}(w,w')
&=
\frac{\theta(t-t'-|{\bf x}-{\bf x}'|)}{2\pi L^{2}}
\Bigg[
yy'
\left(
\delta((t-t')^{2}-r^{2})
+\delta((t-t')^{2}-{\tilde r}^{2})
\right)
\notag\\
&
\hspace{4cm}
-
\frac12
\Big(
\theta((t-t')^{2}-r^{2})
-\theta((t-t')^{2}-{\tilde r}^{2})
\Big)
\Bigg],
\label{gr32}
\\
G^{\nu=\pm 1/2}_{R}(w,w')
&=
\frac{yy'\theta(t-t'-|{\bf x}-{\bf x}'|)}{2\pi L^{2}}
\Big(
\delta((t-t')^{2}-r^{2})
-
\delta((t-t')^{2}-{\tilde r}^{2})
\Big),
\label{gr12}
\end{align}
where
\be
r^{2}=|{\bf x}-{\bf x}'|^{2}+(y-y')^{2}, 
\qquad
{\tilde r}^{2}=|{\bf x}-{\bf x}'|^{2}+(y+y')^{2}.
\label{rrt}
\ee

\subsection{Retarded potential for massless particle scattering}
\label{sectionpotential}

As a simple example of localized energy-momentum source, 
let us consider a scattering event of point particles 
\cite{Tolish:2014bka,Tolish:2016ggo}.
The energy-momentum tensor for 
a particle scattering 
which occurs at a spacetime point
$x^{\mu}_{0}=(t_{0},{\bf z}_{0})$ 
is written as
\be
T_{\mu\nu}
=
\sum_{j,{\rm in}}T^{(j)}_{\mu\nu}
+\sum_{n,{\rm in}}T^{(n)}_{\mu\nu}
+\sum_{i,{\rm out}}T^{(i)}_{\mu\nu}
+\sum_{m,{\rm out}}T^{(m)}_{\mu\nu},
\label{emscat}
\ee
where $T^{(j)}_{\mu\nu}$ is the energy-momentum tensor 
for the $j^{th}$ incoming massive particle,
\be
T^{(j)}_{\mu\nu}
=M_{\rm in}^{(j)}u^{(j)}_{\mu}u^{(j)}_{\nu}
\delta^{(3)}({\bf x}-{\bf z}^{(j)}(t))
\frac{d\tau^{(j)}}{dt}
\frac{\theta(t_{0}-t)}{\sqrt{-{\bar g}}},
\quad
u^{(j)}_{\mu}={\bar g}_{\mu\nu}
\frac{dx^{\nu(j)}}{d\tau^{(j)}},
\label{emscat1}
\ee
$T^{(i)}_{\mu\nu}$ is that for 
the $i^{th}$ outgoing massive particle,
\be
T^{(i)}_{\mu\nu}
=M_{\rm out}^{(i)}u^{(i)}_{\mu}u^{(i)}_{\nu}
\delta^{(3)}({\bf x}-{\bf z}^{(i)}(t))
\frac{d\tau^{(i)}}{dt}
\frac{\theta(t- t_{0})}{\sqrt{-{\bar g}}},
\quad
u^{(i)}_{\mu}={\bar g}_{\mu\nu}
\frac{dx^{\nu (i)}}{d\tau^{(i)}},
\label{emscat2}
\ee
$T^{(n)}_{\mu\nu}$ is that for 
the $n^{th}$ incoming massless particle,
\be
T^{(n)}_{\mu\nu}
=k^{(n)}_{\mu}k^{(n)}_{\nu}
\delta^{(3)}({\bf x}-{\bf z}^{(n)}(t))
\frac{d\lambda^{(n)}}{dt}
\frac{\theta(t_{0}-t)}{\sqrt{-{\bar g}}},
\quad
k^{(n)}_{\mu}={\bar g}_{\mu\nu}
\frac{dx^{\nu (n)}}{d\lambda^{(n)}},
\label{emscat3}
\ee
and $T^{(m)}_{\mu\nu}$ is that for 
the $m^{th}$ outgoing massless particle,
\be
T^{(m)}_{\mu\nu}
=k^{(m)}_{\mu}k^{(m)}_{\nu}
\delta^{(3)}({\bf x}-{\bf z}^{(m)}(t))
\frac{d\lambda^{(m)}}{dt}
\frac{\theta(t- t_{0})}{\sqrt{-{\bar g}}},
\quad
k^{(m)}_{\mu}={\bar g}_{\mu\nu}
\frac{dx^{\nu (m)}}{d\lambda^{(m)}}.
\label{emscat4}
\ee
Here $M_{\rm in}^{(j)}$ 
and $M_{\rm out}^{(i)}$ are 
the rest masses of
the incoming and 
the outgoing massive particles, respectively.
 $\tau^{(j,i)}$ denotes the proper time of 
 the massive particles 
 while $\lambda^{(n,m)}$ denotes 
 the affine parameter 
 for the null geodesics.
${\bf z}^{(j,i,n,m)}(t)$ satisfy 
${\bf z}^{(j,i,n,m)}(t=t_{0})={\bf z}_{0}$.
Terms proportional to $\delta(t-t_{0})$ 
in $D_{\mu}T^{\mu}_{\ \nu}=0$
implies the energy-momentum conservation 
at the interaction point $x^{\mu}_{0}$,
\be
\sum_{j}M_{\rm in}^{(j)}u_{\mu}^{(j)}
+\sum_{n}k_{\mu}^{(n)}
=
\sum_{i}M_{\rm out}^{(i)}u_{\mu}^{(i)}
+\sum_{\rm out}k_{\mu}^{(m)}.
\label{emcon}
\ee

\begin{figure}
  \centering
\includegraphics[width=7cm]{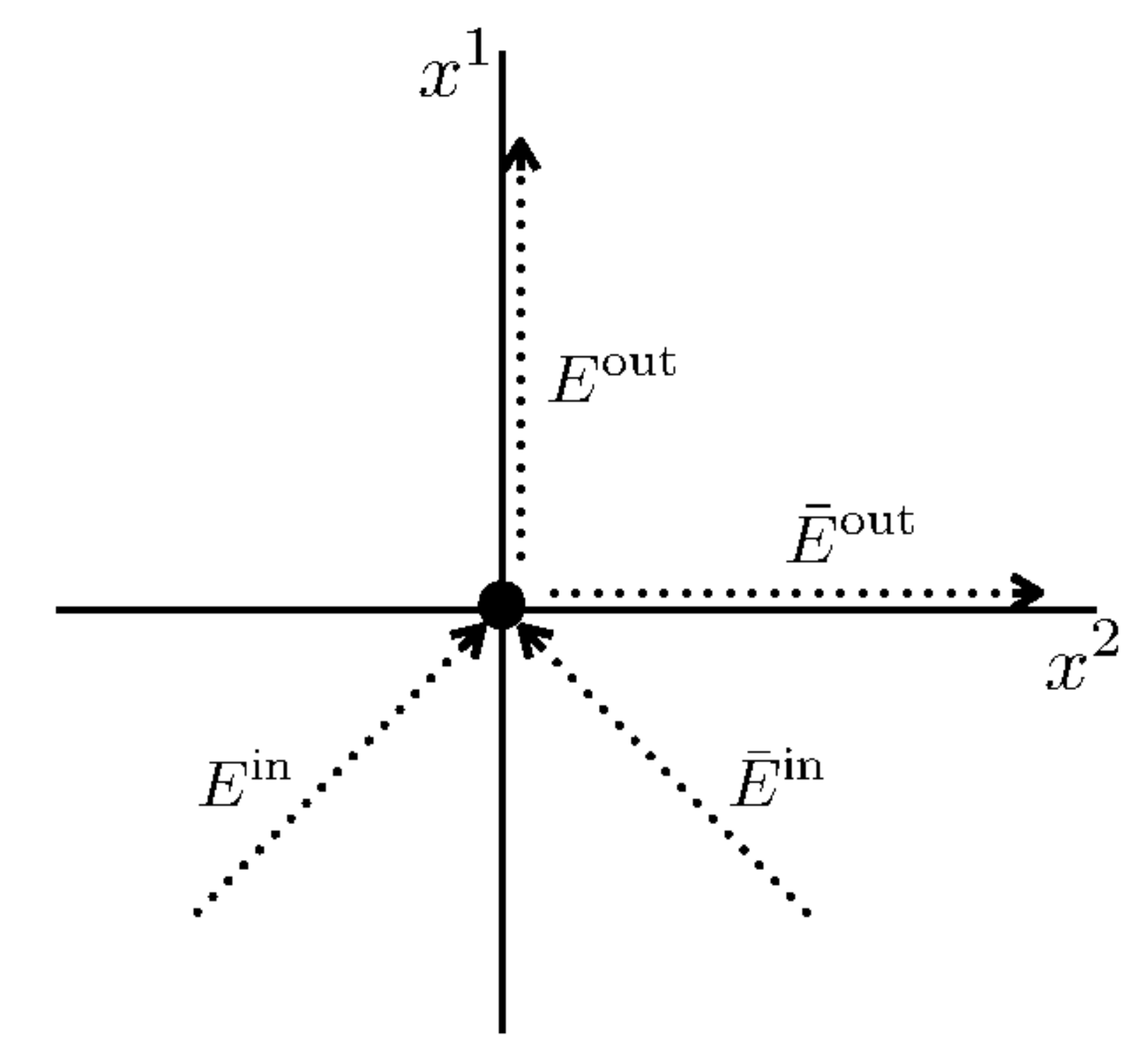}
\caption{A massless scattering  
in the $x^{1}-x^{2}$ plane.}
\label{fig1}
\end{figure}

In order to evaluate the retarded gravitational potential explicitly,
we consider a simple scattering process 
as shown in fig.\ref{fig1}, where the scattering takes place in the
$x^{1}-x^{2}$ plane with a fixed $y=y_{0}$, with two incoming massless particles  
collides at $x^{\mu}_{0}=(t_{0},t_{0},t_{0},y_{0})$
and results in two outgoing massless particles. 
By adopting a suitable Lorentz transformation, we can assume without loss of
generality that the outgoing particles move along the $x^1$ and $x^2$ direction
orthogonally. 
For completeness, we list   some basic materials about the
geodesic motion of point particle 
in AdS space in the Appendix \ref{appgeodesic}.

Note that one can consider more general scattering processes 
in which 
both massless and massive particles are involved. 
In a flat background, one can think of a massive particle at rest
and its decay into a massive and a massless particle. 
It would be one of the most simple examples for 
a particle scattering source 
which gives rise to both ordinary and null memory  
due to the emission of the massive and massless particle 
\cite{Tolish:2014bka}.
In AdS space, evaluation of the retarded potential 
for a decay of a massive particle that is freely moving 
along the geodesic could be done explicitly
but the final expression would be complicated one
and would not be
as illuminating as that for a decay of massive particle 
 at rest in flat space.
 On the other hand, geodesic motion of 
 a massless particle in AdS space 
 are as simple as that in flat space 
 and this is the reason 
 why we focus on a massless scattering source here.
Although we will not consider 
the effect of massive particles
on the retarded potential in AdS space, 
we expect that 
there will be no qualitative differences between 
the behaviour of the retarded potential for 
massless scattering sources and that 
for massive sources,
and the example investigated in this section would suffice 
to capture the characteristics of the retarded potential for 
a particle scattering source.

The energy-momentum tensor for the massless scattering  
is given by
\begin{align}
T_{\mu\nu}(x)
&=
k^{\rm in}_{\mu}k^{\rm in}_{\nu}
\delta(x^{1}-x^{1}_{\rm in})
\delta(x^{2}-x^{2}_{\rm in})
\delta\left(y-y_0\right)
\frac{d\lambda_{\rm in}}{dt}
\frac{\theta(t_{0}-t)}{\sqrt{-{\bar g}}}
\notag\\
& \quad
+
{\bar k}^{\rm in}_{\mu}{\bar k}^{\rm in}_{\nu}
\delta(x^{1}-{\bar x}^{1}_{\rm in})
\delta(x^{2}-{\bar x}^{2}_{\rm in})
\delta\left(y-y_0\right)
\frac{d{\bar \lambda}_{\rm in}}{dt}
\frac{\theta(t_{0}-t)}{\sqrt{-{\bar g}}}
\notag\\
&
\quad
+
k^{\rm out}_{\mu}k^{\rm out}_{\nu}
\delta(t-x^{1})
\delta(x^{2}-t_{0})
\delta(y-y_{0})
\frac{d\lambda_{\rm out}}{dt}
\frac{\theta(t-t_{0})}{\sqrt{-{\bar g}}}
\notag\\
&
\quad
+
{\bar k}^{\rm out}_{\mu}{\bar k}^{\rm out}_{\nu}
\delta(x^{1}-t_{0})
\delta(t-x^{2})
\delta(y-y_{0})
\frac{d{\bar \lambda}_{\rm out}}{dt}
\frac{\theta(t-t_{0})}{\sqrt{-{\bar g}}}
,
\label{emtensorsim}
\end{align}
where we set
\begin{align}
x^{1}_{\rm in}(t) 
&=\frac{1}{\sqrt{2}}(t+c_{1}), 
\quad
x^{2}_{\rm in}(t) 
=\frac{1}{\sqrt{2}}(t+c_{1}), 
\quad
c_{1}=(\sqrt{2}-1)t_{0},
\notag\\
{\bar x}^{1}_{\rm in}(t) 
&=\frac{1}{\sqrt{2}}(t+c_{1}), 
\quad
{\bar x}^{2}_{\rm in}(t) 
=-\frac{1}{\sqrt{2}}(t+c_{2}),
\quad
c_{2}=-(\sqrt{2}+1)t_{0}.
\label{xin}
\end{align} 
Now we are ready to compute the retarded potential from the
  outgoing and incoming massless particles.
Note that $\chi^{a}_{\ a}=\chi_{ya}={\tilde \chi}=0$ 
for the source \eqref{emtensorsim}. 

\parabf{Contribution from outgoing massless particles:}
First we consider the contribution from 
the outgoing massless particles
for which the energy-momentum tensor is
\begin{align}
T^{\rm out}_{\mu\nu}(x)
&=
\frac{y^{2}}{L^{2}}E^{\rm out}  
n^{\rm out}_{\mu}n^{\rm out}_{\nu}
\delta(t-x^{1})
\delta(x^{2}-t_{0})
\delta\left(y-y_0\right)
\theta(t-t_{0})
\notag\\
&
\quad
+
\frac{y^{2}}{L^{2}} {\bar E}^{\rm out}
{\bar n}^{\rm out}_{\mu}{\bar n}^{\rm out}_{\nu}
\delta(x^{1}-t_{0})
\delta(t-x^{2})
\delta(y-y_0)
\theta(t-t_{0}),
\label{emtensorout}
\end{align}
where 
$n^{\rm out}_{\mu}
=-\delta^{0}_{ \mu}+\delta^{1}_{ \mu}$,
${\bar n}^{\rm out}_{\mu}
=-\delta^{0}_{ \mu}+\delta^{2}_{ \mu}$ 
are the unit tangents to the trajectories of 
the outgoing massless particles
and $E^{\rm out},{\bar E}^{\rm out}$ are 
the energies of each of the outgoing massless particles.

The retarded potential is evaluated as
\begin{align}
\chi^{\rm out}_{ab}
&=
\frac{8 G E^{\rm out}}{L^{2}}
 n^{\rm out}_{a}n^{\rm out}_{b}
\int dt'
\theta(t'-t_{0})
\left[
y y_{0}
\Big(
\delta((t-t')^{2}-r^{2})
+\delta((t-t')^{2}-{\tilde r}^{2})
\Big)
\right.
\notag\\
&\hspace{3cm}
\left.
-
\frac12
\Big(
\theta((t-t')^{2}-r^{2})
-\theta((t-t')^{2}-{\tilde r}^{2})
\Big)
\right]_{x'^{1}=t',x'^{2}=t_{0},y'=y_{0}}
\notag\\
&\quad
+
\frac{8 G {\bar E}^{\rm out} }{L^{2}}
{\bar n}^{\rm out}_{a}{\bar n}^{\rm out}_{b}
\int dt'
\theta(t'-t_{0})
\left[
y y_{0}
\Big(
\delta((t-t')^{2}-r^{2})
+\delta((t-t')^{2}-{\tilde r}^{2})
\Big)
\right.
\notag\\
&\hspace{3cm}
\left.
-
\frac12
\Big(
\theta((t-t')^{2}-r^{2})
-\theta((t-t')^{2}-{\tilde r}^{2})
\Big)
\right]_{x'^{1}=t_{0},x'^{2}=t',y'=y_{0}}
\notag\\
&=
\frac{2G}{L^{2}}
\left(
\frac{E^{\rm out}n^{\rm out}_{a}n^{\rm out}_{b}}{t-x^{1}}
+
\frac{{\bar E}^{\rm out}
{\bar n}^{\rm out}_{a}{\bar n}^{\rm out}_{b}}
{t-x^{2}}
\right)
\left[(2yy_{0}-U)\theta(u)
+(2yy_{0}+{\tilde U})\theta({\tilde u})
\right]
,
\label{retms2}
\end{align}
where we have used 
$\theta(U)=\theta(u)$ and 
$\theta({\tilde U})=\theta({\tilde u})$ with
\begin{align}
u&=t-t_{0}-r_{0},
\quad
U=(t-t_{0})^{2}-r^{2}_{0},
\quad 
r_{0}^{2}
=(x^{1}-t_{0})^{2}+(x^{2}-t_{0})^{2}+(y-y_{0})^{2},
\\
{\tilde u}&=t-t_{0}-{\tilde r}_{0},
\quad
{\tilde U}=(t-t_{0})^{2}-{\tilde r}_{0}^{2}, 
\quad
{\tilde r}_{0}^{2}
=(x^{1}-t_{0})^{2}+(x^{2}-t_{0})^{2}+(y+y_{0})^{2}.
\label{UUt}
\end{align}


\parabf{Contribution from incoming massless particles:}

Next we consider the retarded potential 
due to the incoming massless particles.
The energy-momentum tensor is
\begin{align}
T^{\rm in}_{\mu\nu}(x)
&=
\frac{y^{2}}{L^{2}} E^{\rm in} 
m^{\rm in}_{\mu}m^{\rm in}_{\nu}
\delta(x^{1}-x^{1}_{\rm in})
\delta(x^{2}-x^{2}_{\rm in})
\delta\left(y-y_0\right)
\theta(t_{0}-t)
\notag\\
&
\quad
+
\frac{y^{2}}{L^{2}} {\bar E}^{\rm in}
{\bar m}^{\rm in}_{\mu}{\bar m}^{\rm in}_{\nu}
\delta(x^{1}-{\bar x}^{1}_{\rm in})
\delta(x^{2}-{\bar x}^{2}_{\rm in})
\delta(y-y_0)
\theta(t_{0}-t),
\label{emtensorin}
\end{align}
where 
$m^{\rm in}_{\mu}=
-\delta^{0}_{ \mu}
+(\delta^{1}_{ \mu}+\delta^{2}_{ \mu})/\sqrt{2}$, 
${\bar m}^{\rm in}_{\mu}=
-\delta^{0}_{ \mu}
+(\delta^{1}_{ \mu}-\delta^{2}_{ \mu})/\sqrt{2}$
are the unit tangents to the trajectories of 
the incoming massless particles.
The energy-momentum conservation \eqref{emcon} 
at $t=t_{0}$ is
solved by
\be
E^{\rm in}=E^{\rm out}, 
\qquad 
{\bar E}^{\rm in}
={\bar E}^{\rm out}=(\sqrt{2}-1)E^{\rm in}.
\label{emcon2}
\ee
The retarded potential 
due to the incoming massless particles 
is given by
\begin{align}
\chi^{\rm in}_{ab}
&=
\frac{4Gyy_{0}}{L^{2}}
\left(
\frac{E^{\rm in}m^{\rm in}_{a}m^{\rm in}_{b}}
{t+c_{1}-\frac{x^{1}+x^{2}}{\sqrt{2}}}
+
\frac{{\bar E}^{\rm in}
{\bar m}^{\rm in}_{a}{\bar m}^{\rm in}_{b}}
{t-t_{0}-\frac{x^{1}-x^{2}}{\sqrt{2}}}
\right)
\left(\theta(-u)+\theta(-{\tilde u})\right)
\notag\\
&\quad
-
\frac{4G}{L^{2}}
\left[
E^{\rm in}m^{\rm in}_{a}m^{\rm in}_{b}
\left(
\theta(-u)\int^{t_{-}}_{t_{c}} dt'
-
\theta(-{\tilde u})
\int^{{t}_{+}}_{t_{c}} dt'
+
(\theta(u)-\theta({\tilde u}))
\int^{t_{0}}_{t_{c}} dt'
\right)
\right.
\notag\\
&\quad
\left.
+
{\bar E}^{\rm in}
{\bar m}^{\rm in}_{a}{\bar m}^{\rm in}_{b}
\left(
\theta(-u)\int^{{\bar t}_{-}}_{t_{c}} dt'
-
\theta(-{\tilde u})
\int^{{{\bar t}}_{+}}_{t_{c}} dt'
+
(\theta(u)-\theta({\tilde u}))
\int^{t_{0}}_{t_{c}} dt'
\right)
\right]
\notag\\
&
=
\frac{4Gyy_{0}}{L^{2}}
\left(
\frac{E^{\rm in}
m^{\rm in}_{a}m^{\rm in}_{b}}
{t+c_{1}-\frac{x^{1}+x^{2}}{\sqrt{2}}}
+
\frac{{\bar E}^{\rm in}
{\bar m}^{\rm in}_{a}{\bar m}^{\rm in}_{b}}
{t-t_{0}-\frac{x^{1}-x^{2}}{\sqrt{2}}}
\right)
\left(\theta(-u)+\theta(-{\tilde u})\right)
\notag\\
&\quad
-
\frac{4G}{L^{2}}
\left[
E^{\rm in}m^{\rm in}_{a}m^{\rm in}_{b}
\Big(
\theta(-u)t_{-}
-
\theta(-{\tilde u}){t}_{+}
+
(\theta(u)-\theta({\tilde u}))t_{0}
\Big)
\right.
\notag\\
&\qquad\qquad
\left.
+
{\bar E}^{\rm in}
{\bar m}^{\rm in}_{a}{\bar m}^{\rm in}_{b}
\left(
\theta(-u){\bar t}_{-}
-
\theta(-{\tilde u}){{\bar t}}_{+}
+
(\theta(u)-\theta({\tilde u}))t_{0}
\right)
\right],
\notag\\
&=
\frac{2G}{L^{2}}
\left(
\frac{E^{\rm in}
m^{\rm in}_{a}m^{\rm in}_{b}}
{t+c_{1}-\frac{x^{1}+x^{2}}{\sqrt{2}}}
+
\frac{{\bar E}^{\rm in}
{\bar m}^{\rm in}_{a}{\bar m}^{\rm in}_{b}}
{t-t_{0}-\frac{x^{1}-x^{2}}{\sqrt{2}}}
\right)
\left[(2yy_{0}-U)\theta(-u)
+(2yy_{0}+{\tilde U})\theta(-{\tilde u})
\right]
,
\label{retms3}
\end{align}
where $t_{c}$ is an infrared cutoff 
whose dependence cancels out in the end.
$t_{\mp}$ and ${\bar t}_{\mp}$ are 
the solution to 
$(t-t_{-})^{2}-r^{2}(t_{-})=0$, 
$(t-t_{+})^{2}-{\tilde r}^{2}(t_{+})=0$
for the unbarred null geodesic
and 
$(t-{\bar t}_{-})^{2}-r^{2}({\bar t}_{-})=0$, 
$(t-{\bar t}_{+})^{2}-{\tilde r}^{2}({\bar t}_{+})=0$
for the barred null geodesic.
Explicitly, it is
\begin{align}
t_{\mp}
&=
\frac{t^{2}
-x^{1}(x^{1}-\sqrt{2}c_{1})
-x^{2}(x^{2}-\sqrt{2}c_{1})
-(y\mp y_{0})^{2}-c_{1}^{2}}
{2(t+c_{1}-\frac{x^{1}+x^{2}}{\sqrt{2}})},
\notag\\
{\bar t}_{\mp}
&=
\frac{t^{2}-x^{1}(x^{1}-\sqrt{2}c_{1})
-x^{2}(x^{2}-\sqrt{2}c_{2})
-(y\mp y_{0})^{2}-3t_{0}^{2}}
{2(t-t_{0}-\frac{x^{1}-x^{2}}{\sqrt{2}})}.
\label{tr}
\end{align}

\parabf{Behaviour of the retarded potential:}
Putting together \eqref{retms2} and \eqref{retms3} 
and making use of the identity
\begin{align}
\left[(2yy_{0}-U)\theta(u)
+(2yy_{0}+{\tilde U})\theta({\tilde u})
\right]
+
\left[(2yy_{0}-U)\theta(-u)
+(2yy_{0}+{\tilde U})\theta(-{\tilde u})
\right]
=0,
\label{sumzero}
\end{align}
the retarded potential for the massless scattering source 
$\chi=\chi^{\rm out}+\chi^{\rm in}$
is obtained as 
\be
\chi_{ab}
=
\frac{2G}{L^{2}}
\left(
\alpha_{ab}-\beta_{ab}
\right)
\left[(2yy_{0}-U)\theta(u)
+(2yy_{0}+{\tilde U})\theta({\tilde u})
\right],
\quad
\chi_{ya}=0, 
\quad
{\tilde \chi}=0,
 \label{wholeret}
\ee
where we have defined $\alpha_{ab}$ and $\beta_{ab}$ by
\begin{align}
\alpha_{ab}
:=
\frac{E^{\rm out}n^{\rm out}_{a}n^{\rm out}_{b}}
{t-x^{1}}
+
\frac{{\bar E}^{\rm out}{\bar n}^{\rm out}_{a}{\bar n}^{\rm out}_{b}}
{t-x^{2}},
\qquad
\beta_{ab}
:=
\frac{E^{\rm in}m^{\rm in}_{a}m^{\rm in}_{b}}
{t+c_{1}-\frac{x^{1}+x^{2}}{\sqrt{2}}}
+
\frac{{\bar E}^{\rm in}{\bar m}^{\rm in}_{a}{\bar m}^{\rm in}_{b}}
{t-t_{0}-\frac{x^{1}-x^{2}}{\sqrt{2}}}.
\label{alphabeta}
\end{align}
The result \eqref{wholeret} gives
the original metric perturbation $\gamma_{\mu\nu}$
\footnote{
The gauge condition \eqref{gauge} in this case reads 
$\partial_{a}\gamma^{a}_{\ b}=0$.
This implies
$(\alpha_{ab}-\beta_{ab})K^{a}\delta(u)=0$
and
$(\alpha_{ab}-\beta_{ab})
{\tilde K}^{a}\delta({\tilde u})=0$, 
respectively,
where
 $K^{a}=\partial^{a}u=-(t^{a}+r_{0}^{a})$ 
 and 
 ${\tilde K}^{a}=\partial^{a}{\tilde u}
 =-(t^{a}+{\tilde r}_{0}^{a})$.
}:
\be
\gamma_{ab}=\frac{L^{2}}{y^{2}}\chi_{ab},\qquad 
\gamma_{ya}=0,\qquad \gamma_{yy}=0.
\label{chitogamma2}
\ee 
We note that in \eqref{wholeret}, the terms proportional 
to $U$ and ${\tilde U}$ are the tail contribution.
The ratio of the direct and tail contributions 
of the gravitational waves which directly come from
the source event is thus
\begin{align}
\frac{\text{tail}}{\text{direct}}
\approx
\frac{r_{0} \Delta t}{y y_{0}},
\end{align}
where $\Delta t :=t-t_{0}-r_{0}$ 
measures the time passed since 
the passage of gravitational waves.

\begin{figure}
\centering
  \includegraphics[width=7cm]{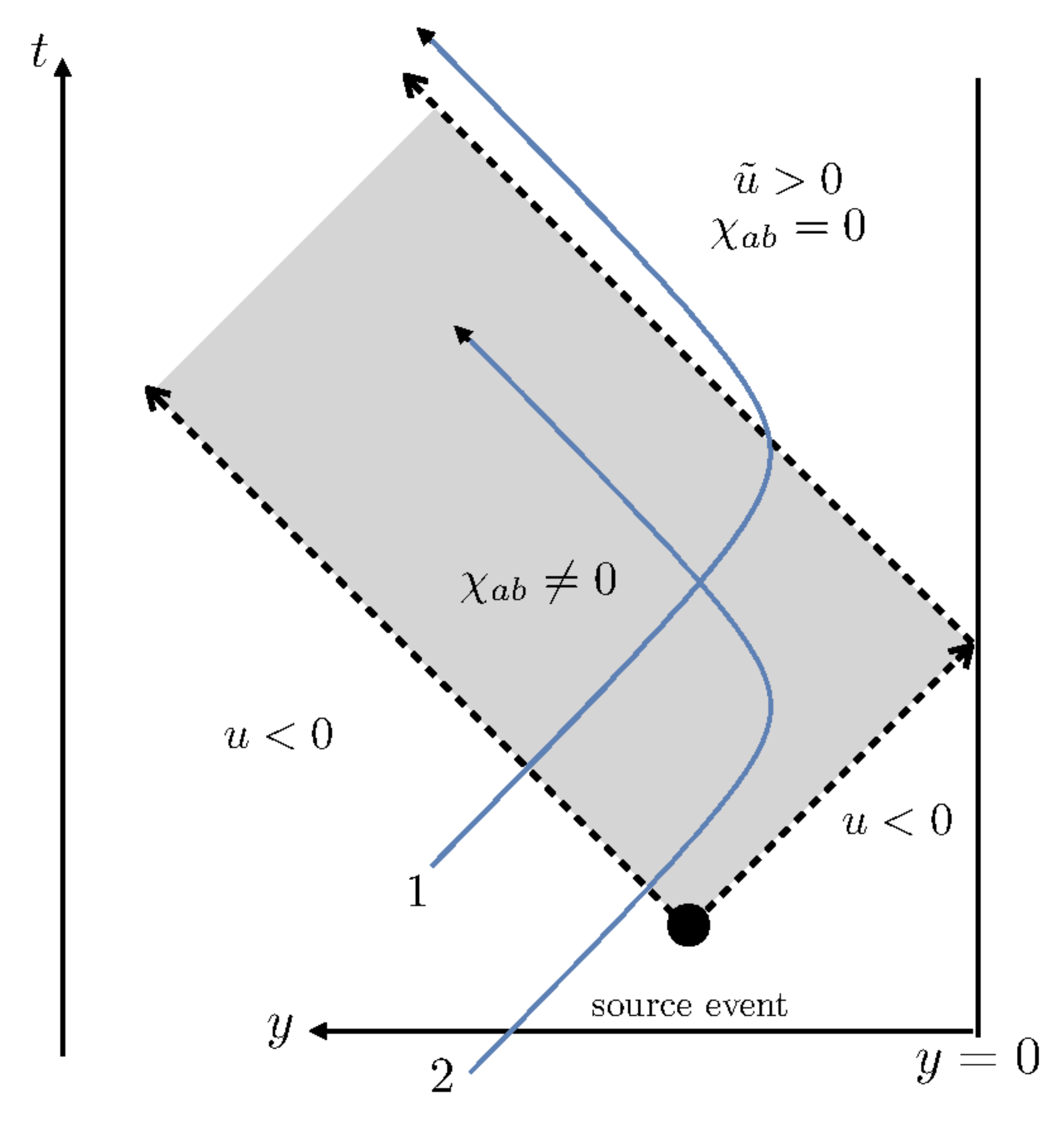}
\caption{Propagation of the gravitational waves
and timelike geodesics.}
\label{fig2}
\end{figure}
Fig.2 shows a 2-dimensional spacetime diagram 
for the behaviour of the retarded potential $\chi_{ab}$ 
where the $x^{i}$ directions are suppressed. 
The bullet represents the source event, 
the dotted lines represent the propagation of gravitational waves 
 traveling at the speed of light. 
Gravitational waves and its tail reach the AdS boundary 
in a finite time 
and then will get reflected back to the bulk.

Let us consider two timelike observers whose geodesics
are given by  
\be
y^{2}(\tau_{1,2})=(t(\tau_{1,2})-T_{1,2})^{2}+L^{2},\quad
x^{1}=\text{const.}, \quad x^{2}=\text{const.},
\ee
where $\tau_{1,2}$ are the proper times
and $T_{1,2}$ are constants with $T_{2}<t_{0}<T_{1}$
for the observer 1 and 2, respectively.
The worldlines of these observers are represented by the
solid curves in fig.\ref{fig2}. 
The retarded potential $\chi_{ab}$ for these observers 
acquires a nonzero contribution discontinuously 
on the light cone $t=t_{0}+{r}$ due to the direct contribution.
In the region $t_{0}+{r}< t < t_{0}+{\tilde r}$,
 the tail contribution is nonzero and  reduce 
 the effect of the potential $\chi_{ab}$. 
When $t\approx t_{r}+{yy_{0}}/{r_{0}}$, 
the tail contribution 
becomes comparable to the direct contribution.
Once $t=t_{0}+{\tilde r}$, that happens to the observer 1,
 then $\chi_{ab}$ vanishes discontinuously 
as the second term in \eqref{retms2} 
which is proportional to $\theta({\tilde u})$
comes into action. 
This means that the contribution from 
the reflected gravitational waves 
cancels exactly that from those directly from the source
and the spacetime goes back to the original vacuum AdS space. 
It follows that $\chi_{ab}$ is nonzero only 
in the gray region in between the two light cones.
Note that for the observer 2, $t\geq t_{0}+{\tilde r}$ will
never be realized and the cancellation does not occur.

\section{Memory Effect in AdS Space}
\label{sectionmemory}

In flat space, the memory effect is defined as 
a permanent displacement in the geodesic deviation of a pair
of test particles after the passage of gravitational waves.
In this paper,
we will use this observable to give 
a definition for the memory effect in AdS space.
The geodesic deviation equation is
\be
u^{\rho}\nabla_{\rho}(u^{\sigma}\nabla_{\sigma}D^{\mu})
=
-R^{\mu}_{\ \alpha\beta\gamma}u^{\alpha}D^{\beta}u^{\gamma}.
\label{gde}
\ee
Here we consider two nearby test particles initially at rest 
as a gravitational wave detector.
$D^{\mu}$ is the deviation vector of the test particles 
and $u^{\mu}={dx^{\mu}}/{d\tau}$ is the unit tangent to the geodesic 
of one of the test particles. 
In what follows, we will call this geodesic 
the central geodesic $\gamma$ 
which is, at the zeroth order of the perturbation,
 parametrized by the proper time $\tau$ 
 ($-\pi L/2 < \tau < \pi L/2$) as 
\be
y(\tau)=
\frac{L}{\cos \left(\frac{\tau}{L}\right)}, 
\qquad
t(\tau)=L\tan \left(\frac{\tau}{L}\right),
\qquad
x^{1}=x^{2}=\text{const.},
\label{cgeodesic}
\ee
with $y^{2}(\tau)=t^{2}(\tau)+L^{2}$.

In general, the geodesic deviation
\be
D^\mu = \bar{D}^\m + \d D^\mu 
\ee
gets an intrinsic contribution $\bar{D}^\m$
from the curvature of the background spacetime and this is independent of the
contribution $\d D^\mu$
from the gravitational wave that we are interested in.
If one subtract out the contribution from the
background curvature, one can obtain from \eq{gde} a second order differential
equation for $\d D^\m$.
This
can be solved in principle
and will give generally an integrated expression for $\d D^\m$
in terms of the full history of the metric perturbation.
This is much more complicated than the local expression \eq{memory-flat}
obtained
in the flat case. In the following, we show that a simple local expression can
be obtained if the memory $\d D^\m$  is observed with
respect to a certain local inertial
frame, the Fermi Normal Coordinate, around the central geodesic $\g$.

\subsection{Gravitational radiation induced geodesic derivation}

In order to simplify the study of the geodesic derivation, it is desirable
to adopt a coordinate system that is somehow adapted to the central geodesic.
In this regard, we find it convenient to 
use the
 Fermi Normal Coordinates (FNC) \cite{Manasse:1963zz}
  associated with the central geodesic $\gamma(\tau)$, 
 $x_{F}^{\alpha}=(t_{F}=\tau,x^{r}_{F})$, $r=1,2,3$, 
 where the time coordinate is identified with the proper time 
 and the spatial coordinates $x^{r}_{F}$ are
 parametrized by the affine parameter of geodesics 
 perpendicular to the central geodesic $\gamma(\tau)$
 on which $x_{F}^{\alpha}=(\tau,0,0,0)$.
 In the following, the subscript $F$ stands for quantity in the FNC.
The Fermi Normal Coordinate system is a locally flat coordinate system 
attached to the central geodesic,
i.e. on the central geodesic $\gamma(\tau)$,
the metric is given by Minkowski metric 
$g^{F}_{\mu\nu}=\eta_{\mu\nu}$
and the Christoffel symbols vanish, 
$(\Gamma_{F})^{\mu}_{\alpha\beta}=0$.
In the neighborhood of the central geodesic, 
the metric receives corrections from the spacetime curvature
which begins with the quadratic term in $x_{F}^{r}$  
of the form $(R_{F})_{\mu r \nu s}x_{F}^{r}x_{F}^{s}$.
In AdS space, $|R_{F}|\simeq L^{-2}$ and the use of FNC
at a spacetime point $P$ away from the central geodesic 
is valid as long as $(x_{F}^{r})^{2}(P)\ll L^{2}$.
\footnote
{In dS space, due to the exponential expansion 
of the physical distance between two nearby geodesics, 
the use of FNC for the analysis of geodesic deviation equation
will be invalidated at late times. 
In such a case, instead of FNC, 
one should use the conformal Fermi coordinates
\cite{Dai:2015rda}.
}

In the FNC, the tangent to the geodesic is
$u_{F}^{\mu}={dx_{F}^{\mu}}/{d\tau}=\delta^{\mu}_{0}$.
Then the geodesic deviation equation takes the form
\be
\frac{d^{2}}{d\tau^{2}}
{D}_{F}^{r}
=
-{(R_{F})}^{r}_{\ 0s 0}(\tau)
{D}_{F}^{s}.
\label{gdefnc}
\ee
where 
${(R_{F})}^{\mu}_{\ \alpha\beta \gamma}(\tau)$ 
is the Riemann tensor in the FNC
that is evaluated on $\gamma(\tau)$
and it is related to the Riemann tensor 
in the Poincare coordinates by 
\be
{(R_{F})}^{\mu}_{\ \alpha\beta \gamma}(\tau) 
=
\eta^{\mu\delta}R_{\nu\rho\sigma\lambda}
(e_{\delta})^{\nu}(e_{\alpha})^{\rho}
(e_{\beta})^{\sigma}(e_{\gamma})^{\lambda}.
\label{riemannfnc1}
\ee
Here
$(e_{\delta})^{\nu}$ is a set of the orthonormal tetrads 
which is parallel transported along ${\gamma(\tau)}$ 
and satisfies 
\cite{Poisson:2009pwt}
\be
(e_{\alpha})^{\mu}
=
\left(\frac{\partial x^{\mu}}
{\partial x_{F}^{\alpha}}\right)_{\gamma(\tau)},
\qquad
\frac{d}{d\tau}(e_{\alpha})^{\mu}
+\Gamma^{\mu}_{\ \rho\sigma}
(e_{0})^{\rho}(e_{\alpha})^{\sigma}
=0,
\qquad
g_{\mu\nu}(e_{\alpha})^{\mu}(e_{\beta})^{\nu}
=\eta_{\alpha\beta}
,
\label{tetrads}
\ee
where $(e_{0})^{\mu}$ is taken 
as the tangent to $\gamma(\tau)$,
$(e_{0})^{\mu}=u^{\mu}$.
For the background AdS space, 
the orthonormal tetrads 
on \eqref{cgeodesic} are determined as
\begin{align}
{\bar e}_{\alpha}^{\mu}
&:=
(e_{\alpha})^{\mu}|_{g_{\mu\nu}={\bar g}_{\mu\nu}},
\notag\\
{\bar e}_{0}^{\mu}
&=\frac{y^{2}}{L^{2}}\delta^{\mu}_{0}
+\frac{ty}{L^{2}}\delta^{\mu}_{y},
\quad
{\bar e}_{i}^{\mu}=\frac{y}{L}\delta_{i}^{\mu},
\quad
{\bar e}_{3}^{\mu}=-\frac{ty}{L^{2}}\delta^{\mu}_{0}
-\frac{y^{2}}{L^{2}}\delta^{\mu}_{y}
.
\label{btetrad}
\end{align}
The geodesic and parallel transport equations in
the perturbative AdS background are discussed
in Appendix \ref{apppert}.

At the first order of the perturbation $\chi_{ab}$ 
given by \eqref{wholeret}, we obtain
\begin{align}
{(R_{F})}_{r0s0}
&=
\frac{1}{L^{2}}\eta_{rs}
\notag\\
&\quad
+
\frac{1}{2L^{4}} \delta^{i}_{r}\delta^{j}_{s}
\Big[
y^{4}(2\partial_{0}\partial_{(j}\chi_{i)0}
-\partial_{0}^{2}\chi_{ij}
-\partial_{i}\partial_{j}\chi_{00})
+2ty^{3}(\partial_{y}\partial_{(j}\chi_{i)0}
-\partial_{0}\partial_{y}\chi_{ij})
\notag\\
&\quad\hspace{2cm}
-t^{2}y^{2}\partial_{y}^{2}\chi_{ij}
+y^{3}\eta_{ij}\partial_{y}\chi_{00}
-yL^{2}\partial_{y}\chi_{ij}
\Big]
\notag\\
&\quad
-\frac{1}{2L^{5}}
(\delta^{i}_{r}\delta^{3}_{s}+\delta^{3}_{r}\delta^{i}_{s})
\Big[
y^{3}L^{2}(\partial_{y}\partial_{0}\chi_{i0}
-\partial_{y}\partial_{i}\chi_{00})
+ty^{2}L^{2}(\partial_{y}^{2}\chi_{i0}
- \frac1y \partial_{y}\chi_{0i})
\Big]
\notag\\
&\quad
+\frac{y^{2}}{2L^{2}} \delta^{3}_{r}\delta^{3}_{s}
(-\partial_{y}^{2}\chi_{00}
+ \frac1y\partial_{y}\chi_{00}).
\label{riemannfnc2}
\end{align}
In deriving this, we have used \eqref{lriemann} 
in Appendix \ref{appcurvature}, \eqref{xdec}, \eqref{edec} and
\eqref{perconstraint} in Appendix \ref{apppert}
together with \eqref{riemannfnc1}. 
It is remarkable that the perturbations of the geodesic $\delta x^{\mu}$
and the tetrads $\delta e_{\alpha}^{\mu}$
do not appear in the final result.
Note that the first line in \eqref{riemannfnc2}
arises from 
the background curvature in the FNC
and gives an oscillating solution for 
 the background geodesic deviation.
 Below we will focus on the deviation vector 
 with an initial condition 
 $d{\bar D}_{F}^{\mu}/{d\tau}(\tau_{i})=0$.
 In this case we have
 \be
 {\bar D}^{r}_{F}
 =C^{r}\cos\left(\frac{\tau-\tau_{i}}{L}\right),
 \label{bdevvec}
 \ee
 where $C^{r}$ is the initial separation.
 Note that we have explicitly isolated the geodesic derivation \eq{bdevvec}
 due to the AdS background
 curvature. 

Since the Riemann tensor \eqref{riemannfnc2} 
has a complicated form, it may appear not easy to solve 
the geodesic deviation equation  \eq{gdefnc}
which is a second order differential equation 
in the perturbed AdS spacetime. 
However there is a trick.
Let us introduce a tensor  defined by
$\Omega^{\mu}_{\ \nu} :=\nabla_{\nu}(e_{0})^{\mu}$.
It has been demonstrated recently in \cite{Shore:2018kmt}, 
especially for shock wave metrics of 
Aichelburg-Sexl type \cite{Aichelburg:1970dh},
that the tensor $\Omega^{\mu}_{\ \nu}$
are useful for the investigation of the memory effect.
In terms of $\Omega^{\mu}_{\ \nu}$,
the Riemann tensor can be expressed as
\be
\frac{d}{d\tau}\Omega^{\mu}_{\ \nu}
+\Omega^{\mu}_{\ \lambda}\Omega^{\lambda}_{\ \nu}
=-R^{\mu}_{\ \alpha \nu \beta} (e_{0})^{\alpha}(e_{0})^{\beta}
\label{omegar}
\ee
and the deviation vector $D^{\mu}$ satisfies 
the first order differential equation \cite{Wald:1984rg}
\be
(e_{0})^{\rho}\nabla_{\rho}D^{\mu}
=\Omega^{\mu}_{\ \nu}D^{\nu}.
\label{omegadev}
\ee
In the FNC, we have
\be
\frac{d}{d\tau}(\Omega_{F})^{\mu}_{\ \nu}
+
(\Omega_{F})^{\mu}_{\ \lambda}(\Omega_{F})^{\lambda}_{\ \nu}
=
-(R_{F})^{\mu}_{\ 0 \nu 0}, 
\label{omegarfnc}
\ee
and
\be
\frac{d}{d\tau}D_{F}^{\mu}
=(\Omega_{F})^{\mu}_{\ \nu}D_{F}^{\nu}.
\label{omegadevfnc}
\ee
Thus the introduction of $\Omega^{\mu}_{\ \nu}$ and the employment of FNC
allow us to greatly reduce the problem
of determining $D_\m$ from solving a second order partial differential
equations to solving a first order ordinary differential equation. 

To proceed, let us compute $(\Omega_{F})^{\mu}_{\ \nu}$:
\be
(\Omega_{F})^{\mu}_{\ \nu}
=
\eta^{\mu\sigma}g_{\alpha \lambda}
(e_{\sigma})^{\lambda}(e_{\nu})^{\beta}
\nabla_{\beta}(e_{0})^{\alpha}.
\label{omegafnc1}
\ee
The spatial components of \eqref{omegafnc1} 
for the central geodesic $\gamma(\tau)$ 
in the perturbed AdS background is given by
\begin{align} 
(\Omega_{F})^{r}_{\ s}
&=
({\bar \Omega}_{F})^{r}_{\ s}
+(\delta \Omega_{F})^{r}_{\ s},
\label{omegafnc2}
\end{align} 
where
\begin{align} 
({\bar \Omega}_{F})^{r}_{\ s}
&=
-\frac{t}{L^{2}} \delta^{r}_{\ s},
\notag\\
(\delta \Omega_{F})^{r}_{\ s}
& =
\frac1y 
\left(\frac{t}{L^{2}} \delta x^{y}
-\delta e^{y}_{0}\right)\delta^{r}_{\ s}
+\frac12
\eta^{rp}{\bar g}_{\alpha\lambda}{\bar e}_{p}^{\lambda}
(\partial_{\beta}h^{\alpha}_{\ \gamma}
+\partial_{\gamma}h^{\alpha}_{\ \beta}
-\partial^{\alpha}h_{\beta\gamma})
{\bar e}_{s}^{\beta}{\bar e}_{0}^{\gamma}
\notag\\
&\quad
+
\frac2y \eta^{rp}h_{0\lambda}
{\bar e}_{s}^{y}{\bar e}_{p}^{\lambda}
+
\eta^{rp}{\bar g}_{\alpha\lambda}
(\delta e^{\lambda}_{p}{\bar e}_{s}^{\beta}
\partial_{\beta}{\bar e}_{0}^{\alpha}
+
{\bar e}^{\lambda}_{p}{\bar e}_{s}^{\beta}
\partial_{\beta}{\delta e}_{0}^{\alpha}).
\label{omegafnc3}
\end{align}
where $h_{\mu\nu}=y^{2}L^{-2}\gamma_{\mu\nu}$. 
Putting
$D_{F}^{r}
={\bar D}_{F}^{r}+\delta D_{F}^{r}$ 
into \eqref{omegadevfnc},
the geodesic derivation vector that arises from the perturbation of the AdS metric
satisfies
\begin{align}
\frac{d}{d\tau}
\left(\frac{y}{L}\delta D_{F}^{r}\right)
&=
\frac{y}{L}(\delta \Omega_{F})^{r}_{\ s}
{\bar D}_{F}^{s} + O(\d D_F^2),
\label{omegadevpert}
\end{align}
where ${\bar D}_{F}^{r}$ is given by \eq{bdevvec}.

Now we are to solve the geodesic deviation equation \eqref{omegadevpert}
with the metric perturbation given by the retarded potential
\eqref{wholeret}. 
Consider the specific example of 
\be
x^{1}=x^{2}=t_{0},
\label{x1x2}
\ee
for the central geodesic $\gamma(\tau)$.
Applying \eqref{x1x2} to \eqref{wholeret}, 
we find that $\chi_{0a}=0$
and $f^{\mu}$ given by \eqref{fads} vanishes. 
It then follows from \eqref{pgeodesicsol} 
(with a set of initial conditions 
$\delta x^{\mu}(\tau_{i})=d\delta x^{\mu}/d\tau(\tau_{i})=0$)
that 
$\delta x^{\mu}=\delta e^{\mu}_{0}=0$.
Therefore \eqref{cgeodesic} with \eqref{x1x2}
is indeed a consistent solution to the perturbed geodesic equation.
As a result, we get
\be
(\delta \Omega_{F})^{r}_{\ s}
=\frac12 \frac{d}{d\tau}\chi^{i}_{\ j}
\ \delta^{r}_{\ i}\delta^{j}_{\ s},
\qquad\qquad
(\delta \Omega_{F})^{r}_{\ r}=0
\label{omegars}
\ee
and the geodesic equation \eq{omegadevpert} can be integrated to a closed form
immediately, giving
\be
D_{F}^{i}(\tau)
={\bar D}_{F}^{i}(\tau)
+\frac12 \chi^{i}_{\ j}(\tau)
{\bar D}_{F}^{j}(\tau)
\qquad
(i,j=1,2),
\qquad\qquad
D_{F}^{3}(\tau)
={\bar D}_{F}^{3}(\tau).
\label{gdesolution}
\ee
It is easy to check that \eqref{gdesolution} 
satisfies the geodesic deviation equation \eqref{gdefnc}
in this case. 
The geodesic deviation in the Poincar\'e coordinates 
is given by $D^{\mu}=(e_{\alpha})^{\mu}D_{F}^{\alpha}$.

It is remarkable that the perturbation of the deviation vector \eq{gdesolution}
can be written in terms of the retarded potential directly and takes the
form \eq{memory-flat} exactly as in the flat case. That this is possible is
because
of the adaptation and simplification brought about by the use of the
Fermi Normal
coordinates.


\subsection{Gravitational radiation memory in AdS}

Memory effect as a permanent displacement 
of the geodesic derivation can be defined in the FNC by
\be
\Delta D^{r}_{F}
:=
\delta 
D^{r}_{F}(\tau_f)
-
\delta 
D^{r}_{F}(\tau_i),
\label{memoryfnc}
\ee
where $\tau_{i}$ and $\tau_{f}$ are the proper time
before and after the passage of gravitational waves.

we have ignored the background effect on the geodesic deviation
since we are interested in the geodesic deviation
with gravitational wave origin.
In AdS spacetime, two distinguished types of gravitational wave
detectors can be considered.
One is a detector whose central geodesic 
passes through the region of nonzero retarded potential,
e.g. observer 1 in fig.\ref{fig2}.
The other is a detector whose central geodesic 
stays in the region of nonzero retarded potential,
e.g. observer 2  in fig.\ref{fig2}.
In the former case ($t_{0}<0$), 
we can discuss the memory effect from a viewpoint of 
a vacuum to vacuum transition of the spacetime. 
But now it is obvious that there is 
no memory effect since after the passage 
of the reflected gravitational waves the spacetime 
settles down to the original vacuum AdS space
described by \eqref{ads}
due to the cancellation of the retarded potential.
In the latter case ($t_{0}>0$), 
 the gravitational wave detector 
will always be under the influence of the retarded potential
since the passage of the direct gravitational waves from the source,
and there will be the competition between
the direct and the tail contributions.
From \eqref{wholeret}, the retarded potential 
for the central geodesic \eqref{cgeodesic}
 with $x^{1}=x^{2}=t_{0}$ is 
\begin{align}
\chi_{ij}
&
=
a_{ij}
\frac{2G}{L^{2}}
\frac{E^{\rm out}}{t-t_{0}}
\Big(2yy_{0}
-(t-t_{0})^{2}+(y-y_{0})^{2}
\Big)
\theta(u)
\notag\\
&
=
a_{ij}
\frac{2G}{L^{2}}
\frac{E^{\rm out}}{t-t_{0}}
\Big(2tt_{0}
+L^{2}-t_{0}^{2}+y_{0}^{2}
\Big)
\theta(u),
\end{align}
where 
$
a_{ij}:= \left(1-\sqrt{2}/2\right)
\left(
\delta_{i}^{1}\delta_{j}^{1}
-\delta_{i}^{1}\delta_{j}^{2}
-\delta_{i}^{2}\delta_{j}^{1}
-\delta_{i}^{2}\delta_{j}^{2}
\right).
$
Here we have used 
$y^{2}=t^{2}+L^{2}$
and have dropped 
the term proportional to $\theta({\tilde u})$.
It can be seen that at late times 
$\chi_{ij}$ will approach a constant value
 \be
\chi_{ij} 
\sim
a_{ij}\frac{4Gt_{0}}{L}.
 \ee
Therefore the geodesic deviation 
will also approach some constant value and 
yields a permanent displacement $\Delta D^{r}_{F}\neq 0$
while the metric satisfies the vacuum Einstein equation 
in AdS space.

\subsection{Asymptotic expansion near the AdS boundary}
The asymptotic behaviour of 
the retarded potential near the AdS boundary 
can be obtained by the Taylor expansion of 
\eqref{wholeret} at $y\to 0$.
Using
\begin{align}
(2yy_{0}-U)\theta(U)
&=
-U_{0}\theta(U_{0})
+
\left[
\theta(U_{0})
+U_{0}(\delta(U_{0})
-2y_{0}^{2}\delta'(U_{0}))
\right]
y^{2}
\notag\\
&\quad
+
2y_{0}\left[
\delta(U_{0})
+
U_{0}\left(\delta'(U_{0})
-\frac{2y_{0}^{2}}{3}\delta''(U_{0})\right)
\right]
y^{3}
+{\cal O}(y^{4}),
\label{asymptotic}
\\
(2yy_{0}+{\tilde U})\theta({\tilde U})
&=
U_{0}\theta(U_{0})
-
\left[
\theta(U_{0})
+
U_{0}(\delta(U_{0})-2y_{0}^{2}\delta'(U_{0}))
\right]
y^{2}
\notag\\
&\quad
+
2y_{0}\left[
\delta(U_{0})
+
U_{0}\left(\delta'(U_{0})
-\frac{2y_{0}^{2}}{3}\delta''(U_{0})\right)
\right]
y^{3}
+{\cal O}(y^{4}),
\label{asymptotic2}
\end{align}
where
$ U_{0}:=
(t-t_{0})^{2}-(x^{1}-t_{0})^{2}
-(x^{2}-t_{0})^{2}-y_{0}^{2}$, we obtain
\be
\chi_{ab}
=
\frac{4 G(yy_{0})^{3}}{3L^{2}r_{\rm B}^{3}}
\delta(u_{0})
\left(1
-r_{\rm B}\frac{\partial}{\partial t}\right)
(\alpha_{ab}-\beta_{ab})
+{\cal O}(y^{5}),
\label{retmsas}
\ee
where
$u_{0}=t-t_{0}-r_{\rm B}$ and
\be
r_{\rm B}
=
\sqrt{(x^{1}-t_{0})^{2}+(x^{2}-t_{0})^{2}+y_{0}^{2}}
\ee
is the distance from the scattering event 
to a point under consideration at the boundary.
We note there is a delta function singularity localized at the lightcone
from the source and it arises because of a careful handling of the
step-functions as  distribution.

The vacuum expectation value of boundary energy-momentum tensor   
 is given by the formula \cite{deHaro:2000vlm} 
\be
\langle T^{\rm B}_{ab}\rangle
=\frac{3L^{2}}{16\pi G}\lim_{y\to 0}\frac{1}{y^{3}}
\chi_{ab}
.
\label{bemtensor}
\ee
Substituting \eqref{retmsas} into \eqref{bemtensor}
yields
\be
\langle T^{\rm B}_{ab}\rangle
=
\frac{ y_{0}^{3}}{4\pi r_{\rm B}^{3}}
\delta(u_{0})
\left(1
-r_{\rm B}\frac{\partial}{\partial t}\right)
(\alpha_{ab}-\beta_{ab}).
\label{bemtensor2}
\ee
Thus the boundary energy-momentum tensor 
is localized on the hyperbola $u_{0}=0$.

\subsection{AdS shock wave}
\label{sectionshock}

In addition to displacement memory, gravitational wave may induce other observable
effects on a detector. In the flat case, a notable effect is that a shock
wave would give rises to a velocity memory in the form of a relative velocity kick
between two nearby particles. It is interesting to know if a shock wave in AdS
spacetime would give rises to a velocity memory; and if so, how different is it
would be from the flat case?

Let us consider the shock wave limit
of the retarded potential.
Naively if we take $t_{0}\to -\infty$ 
in \eqref{emtensorsim}, 
the source describes 
massless particles which travel 
at the speed of light forever. 
In flat space, the gravitational field 
of such a massless particle
is described by 
Aichelburg-Sexl shock wave metric 
\cite{Aichelburg:1970dh}
which is localized on the light cone.
In AdS space, such a shock wave metric 
was first obtained in \cite{Hotta:1992qy}. 
Note that, however, the AdS shock wave metric 
cannot be obtained by 
simply taking $t_{0}\to -\infty$ 
in the retarded solution  \eqref{retms2}
since it vanishes for $t_{0}\to -\infty$. 
Instead, following \cite{Voronov:1974jp}, 
we start with a massless source with 
a finite extent $\Delta$ 
in the $x^{1}$ direction 
with the energy-momentum tensor given by 
\begin{align}
T^{\rm sw}_{\mu\nu}(x)
&=
\frac{y^{2}}{L^{2}} E n_{\mu}n_{\nu}
\frac{1}{\Delta}
\left[\theta(t-x^{1})-\theta(t-x^{1}-\Delta)\right]
\delta(x^{2})\delta(y-y_{0})
\theta(x^{1}- t_{0}).
\label{masslesssimplemod}
\end{align}
Here we have picked one of the out going massless particles
moving in the $x^{1}$ direction 
and set the scattering point as 
$(t_{0},t_{0},0,y_{0})$ for simplicity.

The retarded potential is decomposed into 
the direct and tail parts, $\chi^{\rm sw}_{ab}
=\chi^{\rm direct}_{ab}+\chi^{\rm tail}_{ab}$,
\begin{align}\label{retmsmod}
\chi^{\rm direct}_{ab}
&=
4 G E\frac{y y_{0}}{\Delta L^{2}} n_{a}n_{b}
\left[
\theta(u-\Delta)
\int_{\frac{(t-\Delta)^{2}-(x^{1})^{2}-\rho^{2}}
{2(t-\Delta-x^{1})}}
^{\frac{t^{2}-(x^{1})^{2}-\rho^{2}}{2(t-x^{1})}}
\frac{1}{\sqrt{(x^{1}-x'^{1})^{2}+\rho^{2}}}
dx'^{1}
\right.
\notag\\
&
\left.
+
(\theta(u)-\theta(u-\Delta))
\int_{t_{0}}^{\frac{t^{2}-(x^{1})^{2}-\rho^{2}}{2(t-x^{1})}}
\frac{1}{\sqrt{(x^{1}-x'^{1})^{2}+\rho^{2}}}
dx'^{1}
+(\rho \leftrightarrow {\tilde \rho},
\ u \leftrightarrow {\tilde u})
\right],
\end{align}
\begin{align}
\chi^{\rm tail}_{ab}
&=
- \frac{4 G E}{\Delta L^{2}}n_{a}n_{b}
\left[
\theta(u)
\int_{t_{0}}^{\frac{t^{2}-(x^{1})^{2}-\rho^{2}}
{2(t-x^{1})}}
(t-x'^{1}-r)
dx'^{1}
\right.
\notag\\
&
\left.
-
\theta(u-\Delta)
\int_{t_{0}}^{\frac{(t-\Delta)^{2}-(x^{1})^{2}-\rho^{2}}
{2(t-\Delta-x^{1})}}
(t-\Delta-x'^{1}-r)
dx'^{1}
-(\rho\leftrightarrow {\tilde \rho},
\ r \leftrightarrow {\tilde r},
\ u \leftrightarrow {\tilde u})
\right],
\label{retmsmodt}
\end{align}
where
\begin{align}
u
&=
t-t_{0}-\sqrt{(x^{1}-t_{0})^{2}+\rho^{2}},
\qquad
{\tilde u}
=
t-t_{0}-\sqrt{(x^{1}-t_{0})^{2}+{\tilde \rho}^{2}},
\label{uut}
\\
\rho^{2}
&
=(x^{2})^{2}+(y-y_{0})^{2},
\qquad
{\tilde \rho}^{2}
=(x^{2})^{2}+(y+y_{0})^{2}.
\label{rhorhot}
\end{align}
The shock-wave limit should be taken carefully by
first performing the integrations,  then
taking $t_{0}\to -\infty$ 
and finally the limit $\Delta\to 0$. As a result, we obtain
\begin{align}
\chi^{\rm direct}_{ab}
&
\to
4GE\frac{y y_{0}}{ L^{2}} n_{a}n_{b}
\left[
\frac{2}{t-x^{1}}\theta(t-x^{1})
-\ln
\left(
\frac{\rho^{2}{\tilde \rho}^{2}}{4t_{0}^{2}(t-x^{1})^{2}}
\right)
\delta(t-x^{1})
\right]
,
\label{directlimit}
\\
\chi^{\rm tail}_{ab}
&
\to
-4GE\frac{y y_{0}}{ L^{2}} n_{a}n_{b}
\left[
\frac{2}{t-x^{1}}\theta(t-x^{1})
\right.
\notag\\
&\hspace{1cm}
\left.
+
\left(
2-
\ln
\left(
\frac{\rho^{2}{\tilde \rho}^{2}}
{4t_{0}^{2}(t-x^{1})^{2}}
\right)
-
\frac{\rho^{2}+{\tilde \rho}^{2}}{4y y_{0}}
\ln\left(\frac{{\tilde \rho}^{2}}{\rho^{2}}\right)
\right)
\delta(t-x^{1})
\right]
.
\label{taillimit}
\end{align}
Adding together \eqref{directlimit} 
and \eqref{taillimit},
we obtain
\be
\lim_{\substack{t_{0}\to -\infty \\ \Delta\to 0}}
\chi_{ab}^{\rm sw}
=
\frac{8GE}{L^{2}}
n_{a}n_{b}
\left[
-y y_{0}+
\frac{\rho^{2}+{\tilde \rho}^{2}}{8}
\ln\left(\frac{{\tilde \rho}^{2}}{\rho^{2}}\right)
\right]
\delta(t-x^{1}).
\label{shockads4}
\ee
This is precisely the AdS shock wave geometry
obtained in \cite{Hotta:1992qy,Podolsky:1997ri}
for $y_{0}=L$. 
One can check that the AdS shock wave geometry
$g_{\mu\nu}
={\bar g}_{\mu\nu}
+\delta^{a}_{\ \mu}\delta^{b}_{\ \nu}
\frac{L^{2}}{y^{2}}\chi^{\rm sw}_{ab}$
is in fact a solution to 
the full nonlinear Einstein equation
with a source given by a massless particle 
traveling at the speed of light forever.
Thus we have provided an alternative derivation of
the AdS shock wave by taking a limit of the retarded potential. 
We note that it is crucial in our derivation of the shock wave
geometry 
to take into account of the tail term. Otherwise we will not get the correct
result.
The VEV of boundary energy-momentum tensor 
corresponding to \eqref{shockads4} is
\be
\langle T_{ab}^{\rm B}\rangle
=
n_{a}n_{b}
\frac{2 E y_{0}^{3}}{\pi ((x^{2})^{2}+y_{0}^{2})^{2}}
\delta(t-x^{1}),
\ee
and this is localized on the light cone
\cite{Horowitz:1999gf,Gubser:2008pc}.

\subsection{Velocity-memory of AdS shock wave}

It is known  
that passing through the shock wave causes
a jump in the advanced time coordinate
and a refraction of the geodesic
\cite{Penrose:1968ar,Dray:1984ha,Sfetsos:1994xa,Podolsky:2001vu}.
It is instructive to compare the memory effect of shock wave in AdS spacetime
and flat spacetime. In flat space, the shock wave
induces a permanent displacement 
in the relative velocity of two nearby timelike geodesics
\cite{Tolish:2014bka,Shore:2018kmt,Steinbauer:1997dw,Zhang:2017jma}.
We will show now that the AdS shock wave induces much richer features in the
velocity memory: the relative velocity kick \eqref{relativekick}
in AdS has 
a {\it kink} $u\theta(u)$, a {\it jump} $\theta(u)$ 
and a {\it pulse} $\delta(u)$ in its $v$ component;
and a kink and a jump in 
the $u$, $x^{2}$ and $y$ components.

To start with, let us review the analysis of \cite{Podolsky:2001vu}
for  geodesic motion 
in the AdS shock wave background. 
The shock wave metric is written using a 5 dimensional formalism as
\be
ds^{2}
=
H(Z_{2},Z_{3},Z_{4})\delta(U)dU^{2}
-2dUdV+dZ_{2}^{2}+dZ_{3}^{2}-dZ_{4}^{2}
\label{5dshock}
\ee
where
$
U=(Z_{0}+Z_{1})/\sqrt{2},\ 
V=(Z_{0}-Z_{1})/\sqrt{2}
$
and the 5 dimensional coordinates are subject to a constraint
\be
-2UV+Z_{2}^{2}+Z_{3}^{2}-Z_{4}^{2}=-L^{2}.
\label{5dconst}
\ee
In this description, the shock wave is localized at $U=0$.
In the following, we assume that 
$H$ is a function of $Z_{4}$ only $H=H(Z_{4})$
so that it corresponds to the Hotta-Tanaka AdS shock wave 
as \cite{Podolsky:1997ri}
\be
H(Z_{4})
=\frac{4\sqrt{2}GE}{L}
\left[-2L
+Z_{4}\log
\left(\frac{L+Z_{4}}{L-Z_{4}}\right)
\right].
\ee
The most general solution to the geodesic equation 
in the AdS shock wave background \eqref{5dshock} is 
given by eq.(39) of \cite{Podolsky:2001vu} with $e=-1$,
where it was also pointed out that 
one can always achieve the vanishing velocities
in front of the shock wave,
${\dot Z}^{0}_{p}=dZ_{p}/d\tau(U=0^{-})
=0$, $p=2,3,4$,
by utilizing the symmetry of  the shock wave background.
Thus without loss of generality, one can reduce 
the solution to the form
(eq.(37) of \cite{Podolsky:2001vu}):
\begin{align}
&
U
=L{\dot U}^{0}\sin\left(\frac{\tau}{L}\right),
\qquad
Z_{p}(U)
=
Z_{p}^{0}\sqrt{1-(L{\dot U}^{0})^{-2}U^{2}}
+A_{p}U\theta(U),
\notag\\
&
V(U)
=
\frac12({\dot U}^{0})^{-2}U
+B\theta(U)\sqrt{1-(L{\dot U}^{0})^{-2}U^{2}}
+CU\theta(U),
\label{geosol5d}
\end{align}
where $\tau$ is the proper time of a timelike geodesic 
and
\begin{align}
&A_{4}
=\frac{1}{2}
\left(-\partial_{4}H(0)
+\frac{1}{L^{2}}Z^{0}_{4}
G(0)
\right), \qquad
A_{i}
=\frac{1}{2L^{2}}Z^{0}_{i}
G(0), 
\quad
i =2,3 
\notag\\
&
B
=\frac12 H(0),
\qquad
C
=\frac{1}{8}
\left(-(\partial_{4}H(0))^{2}
-\frac{1}{L^{2}}H(0)^{2}
+(Z^{0}_{4}\partial_{4}H(0))^{2}
\right),
\nn\\
& H(0):=H(Z_{4}^{0}),
\qquad
G(0):= Z^{0}_{4}\partial_{4}H(0)-H(0).
\label{5dcoefficients}
\end{align}
The solution is specified by the constants 
${\dot U}^{0}:=dU/d\tau(U=0^{-})$ 
and $Z^{0}_{p}:=Z_{p}(U=0^{-})$.
We are interested in the geodesic deviation of
two nearby timelike geodesics, $x_{A}(\tau)$ and $x_{B}(\tau)$.
Before crossing the shock wave at $U=0$,
the relative velocity of the two geodesics 
is generally nonzero and it would not be possible to achieve
${\dot Z}^{0}_{p}=0$ for both of the geodesics
by utilizing the symmetry transformations of the coordinates.
However, if the two geodesics are parallel to each other
and are 
at the same velocity (the relative velocity is zero) initially,
then it is possible
to achieve ${\dot Z}^{0}_{p}=0$ 
for both $x_{A}(\tau)$ and $x_{B}(\tau)$
by using the symmetry transformations. 
In what follows we will restrict ourselves to 
this case since it is a very natural
situation to have 
a pair of test particles which 
are initially at rest at the time the shock wave arrive, and 
to consider the relative velocity kick 
induced by the shock wave.
As we will see later, in this case
the geodesic deviation 
in the Poincar\'e coordinates 
can be characterized by $\xi_{0}$, 
the relative separation in $x^{2}$ direction
at $U=0_{-}$.

Let us express the above solution \eqref{geosol5d} 
in the Poincar\'e coordinates ($t,x^{1},x^{2},y$) by
using the coordinate transformation:
\begin{align}
Z_{0}
&=\frac{L}{y}t,
\quad
Z_{1}=-\frac{L}{y}x^{1},
\quad
Z_{2}=\frac{L}{y}x^{2},
\quad
Z_{3}=
\frac{L}{2yy_{s}}
(y^{2}+(x^{1})^{2}+(x^{2})^{2}-t^{2}-y_{s}^{2}),
\notag\\
Z_{4}
&=
\frac{L}{2yy_{s}}
(y^{2}+(x^{1})^{2}+(x^{2})^{2}-t^{2}+y_{s}^{2}),
\label{Ztox}
\end{align}
where $y_{s}$ denotes the value of $y$ coordinate 
at the source massless particle.
The shock wave metric \eqref{5dshock} can then be written as
\be
ds^{2}
=
\frac{L^{2}}{y^{2}}
\left(
\frac{y}{L}H(x^{2},y)\delta(u)du^{2}
-2dudv+(dx^{2})^{2}+dy^{2}
\right),
\label{4dshockmetric}
\ee
where
$u=(t-x^{1})/\sqrt{2}$, $v=(t+x^{1})/\sqrt{2}$
and
\be
H(x^{2},y)
=
4\sqrt{2}GE
\left[
-2
+
\frac{(x^{2})^{2}+y^{2}+y_{s}^{2}}{2yy_{s}}
\log
\left(
\frac{(x^{2})^{2}+(y+y_{s})^{2}}
{(x^{2})^{2}+(y-y_{s})^{2}}
\right)
\right].
\ee
Here $t=x^{1}$ has been imposed by $\delta(u)$.
Using \eqref{Ztox},
the solution \eqref{geosol5d} can be written in terms of
($u,v,x^{2},y$) as
\begin{align}
&y(U)
=
\frac{y_{s}L}{Z_{4}-Z_{3}}
=
\frac{y_{s}L}
{y_{s}\sqrt{1-2U^{2}L^{-2}}+(A_{4}-A_{3})U\theta(U)}
\notag\\
&
u(U)
=\frac{1}{\sqrt{2}}\frac{Z_{0}+Z_{1}}{Z_{4}-Z_{3}}
=\frac{y}{L}U,
\notag\\
&
v(U)
=\frac{1}{\sqrt{2}}\frac{Z_{0}-Z_{1}}{Z_{4}-Z_{3}}
=\frac{y}{L}
\left[
U
+
\Big(B\sqrt{1-2U^{2}L^{-2}}
+CU\Big)
\theta(U)
\right],
\notag\\
&
x^{2}(U)
=
\frac{y_{s}Z_{2}}{Z_{4}-Z_{3}}
=\frac{y}{L}
\left(Z^{0}_{2}\sqrt{1-2U^{2}L^{-2}}
+A_{2}U\theta(U)
\right),
\label{4dshockgeodesic}
\end{align}
where we have set 
${\dot U}^{0}=1/\sqrt{2}$ for simplicity.
We immediately find that 
$y(U=0^{-})=L$ 
and then it follows that
$x^{2}_{0}:= x^{2}(U=0^{-})=Z_{2}^{0}$, 
$Z_{3}^{0}=((x^{2}_{0})^{2}+L^{2}-y_{s}^{2})/(2y_{s})$
and 
$Z_{4}^{0}=((x^{2}_{0})^{2}+L^{2}+y_{s}^{2})/(2y_{s})$.
Hence the solution \eqref{4dshockgeodesic} 
is specified
entirely in terms of $x^{2}_{0}$.

Let us now consider two nearby timelike geodesics
denoted by $x_{A}^{\mu}(\tau)$ and $x_{B}^{\mu}(\tau)$
and the geodesic deviation between them.
As mentioned in the above, 
we consider the case in which
$x_{A}^{\mu}(\tau)$ and 
$x_{B}^{\mu}(\tau)$ are parallel 
with the same velocity. 
Then, let $x_{A}^{\mu}(\tau)$
be specified by \eqref{4dshockgeodesic}
with $(x_{A}^{2})_{0}=x^{2}_{0}$
and
$x_{B}^{\mu}(\tau)$ be that specified by 
$(x_{B}^{2})_{0} = x^{2}_{0}+\xi_{0}$.
$x_{A}(\tau)$ passes the AdS shock wave 
at $\tau=0$, $x^{2}=x^{2}_{0}$ while
 $x_{A}(\tau)$ passes it 
at $\tau=0$, $x^{2}=x^{2}_{0}+\xi_{0}$.
The geodesic deviation is written as
\be
D^{\mu}=x_{B}^{\mu}(\tau)-x_{A}^{\mu}(\tau),
\label{devAB}
\ee
with 
\be
D^{\mu}(U<0)
=\xi_{0}\delta^{\mu}_{\ 2},
\qquad
\frac{dD^{\mu}}{dU}(U<0)=0.
\ee
At the lowest order of $\xi_{0}$, 
we obtain
\begin{align}
&
D^{u}
=
\frac{D^{y}}{L} U,
\qquad
D^{y}
=
-
\frac{y_{s}L\partial_{x^{2}_{0}}(A_{4}-A_{3})\xi_{0}U\theta(U)}
{(y_{s}\sqrt{1-2U^{2}L^{-2}}+(A_{4}-A_{3})U\theta(U))^{2}},
\notag\\
&
D^{v}
=
\frac{D^{y}}{L}
\left[U+
\Big(B\sqrt{1-2U^{2}L^{-2}}+CU
\Big)
\theta(U)
\right]
+\frac{y\xi_{0}}{L}
\partial_{x^{2}_{0}}
\left(B\sqrt{1-2U^{2}L^{-2}}
+C U
\right)\theta(U),
\notag\\
&
D^{2}
=
\frac{D^{y}}{L}
\left(x^{2}_{0}\sqrt{1-2U^{2}L^{-2}}+A_{2}U\theta(U)
\right)
+\frac{y}{L}\xi_{0}
\left(\sqrt{1-2U^{2}L^{-2}}
+\partial_{x^{2}_{0}}A_{2}U\theta(U)
\right).
\label{4dshockdeviation}
\end{align}
The relative velocity kick can be written as
\be
\Delta v^{\mu}
:=\frac{dD^{\mu}}{dU}(U>0)-\frac{dD^{\mu}}{dU}(U<0).
\label{relativekick}
\ee
Using the same terminology as in \cite{Steinbauer:1997dw},
we conclude that $\Delta v^{\mu}$
receives 
a  kink $U\theta(U)$, a jump $\theta(U)$ 
and a pulse $\delta(U)$ 
in the $v$-component,
and  a kink and a jump in 
the $(u, x^{2}, y)$-directions.

We remark that it is possible to introduce FNC to investigate
the geodesic deviation in the AdS shock wave background.
In that case it is expected from 
the flat space results \cite{Shore:2018kmt}
\footnote
{
The flat space result
\be
D^{u}= D^{y}=0, \quad
D^{v}= -\xi_{0}
\left(
\frac{4\sqrt{2}GE}{x_{0}^{2}}\theta(u)
+\frac{32(GE)^{2}}{(x^{2}_{0})^{3}}u\theta(u)
\right),
\quad
D^{2}= \xi_{0}
\left(1
+\frac{4\sqrt{2}GE}{(x^{2}_{0})^{2}}u\theta(u)
\right),
\label{devflatlimit}
\ee
can be obtained by taking 
$y\to L$, $y_{s}\to L$, $L\to \infty$ 
in \eqref{4dshockdeviation}.
}
that the geodesic deviation \eqref{4dshockdeviation} 
in the FNC is nonzero only in the
transverse $x_F^{2}$ and
$y_F$ directions
and so is the relative velocity kick. 
We also note that in the FNC the geodesic
derivation
will not be
simply
multiplied by the overall scale factor $y/L$,
and so it can be inferred from the forth equations of 
\eqref{4dshockgeodesic} and \eqref{4dshockdeviation}
that we will have a permanent displacement in 
the relative velocity kick in the $x^{2}$-direction.


\section{Conclusion and Discussions}
\label{sectionsummary}

We have investigated
the retarded potential and memory effect 
for the particle scattering source in AdS space.
Apart from the tail term, 
the retarded propagator receives  contribution 
from the reflected gravitational waves. 
We evaluated the retarded potential 
for a particle scattering source and found that
the retarded potential contains two kinds of step functions
$\theta(u)$ and $\theta({\tilde u})$ corresponding to
the two types of the gravitational waves,
one came directly from the source and the other experienced a reflection.
As a consequence, the retarded potential is nonzero only 
in a finite domain of the spacetime as shown in fig.\ref{fig2}. 
Once the two contributions become active, they cancel 
each other out giving a vanishing retarded potential
and the spacetime goes back to 
the original vacuum AdS space.
This is a somewhat surprising result.

We have solved the geodesic deviation equation
in the perturbed AdS space by making use of Fermi normal coordinates (FNC) and 
the tensor $\Omega^{\mu}_{\ \nu}$.
We find that in the FNC, the geodesic perturbation vector and the displacement
memory depends linearly and locally on the retarded potential
exactly the same way as in
the flat space \eqref{memory-flat}.
This is a  nice result of this work. 

Even though it
may be expected from the behaviour of 
the retarded potential,
we made it clear that there will be no memory  
for the gravitational wave detector
 which passes through the region of nonzero retarded potential.
 On the other hand, for a detector which stays in the region of
 nonzero retarded potential,
the direct and tail contributions together are shown to 
approach to some constant value at the late time, while the perturbed
metric still satisfies the vacuum Einstein equation. This 
corresponds precisely to a non-vanishing
memory induced by the gravitational radiation.

It is known that the Fefferman-Graham expansion of 
the metric near the AdS boundary ($y=0$) of an asymptotically AdS space 
begins with ${\cal O}(y^{3})$ term and it
does not have terms linear or quadratic in $y$.
Actually we saw that the asymptotic expansion of 
the retarded potential for a particle scattering source
also begins with ${\cal O}(y^{3})$ term as 
the ${\cal O}(y^{0})$ and ${\cal O}(y^{2})$ terms
cancels out between the contributions of the gravitational wave 
with its reflection. It is interesting to understand better if and
how the asymptotic form of the memory is related to the asymptotic symmetries
of AdS space. 

We have considered the memory effect of a shock wave in AdS. We find for a pair
of particles initially at rest relative to each other, the passage of the shock
wave will induce a velocity kick in the relative velocity. Unlike in flat case
where the velocity kick is in the form of a jump $\th(u)$ and a pulse $ \d(u)$,
there is also a kink $u \th(u)$ contribution in the AdS case.

In flat spacetime, the gravitational memory effect from localized
particle source
is characterized \cite{Tolish:2014bka}
by a discontinuity in the retarded potential 
and 
a first order derivative of the delta function in the Riemann tensor.
In the present AdS case, the retarded potential in AdS space get additional
complications compared to 
that in the flat space: there is the multiplication of 
the warp factor $yy_{0}/L^{2}$ and there is also 
an additional contribution from the reflected gravitational waves
proportional to $\theta({\tilde u})$.

In this work, we have  demonstrated that the use of the
Fermi normal coordinates allows us to disentangle in the geodesic
deviation
the background curvature contribution from the gravitational wave contribution,
and
extract the gravitational memory of interest. For general curved spacetime,
our analysis suggests that the use of a certain adapted coordinate system
could be very helpful in allowing
one to dissect the geodesic deviation of test particles and
extract the relevant memory due to gravitational radiation. It is interesting
to understand what properties are needed for the right  local coordinate
system. 
For the deSitter space, due to the background expansion, we find that
the use of conformal Fermi coordinates (CNC) seems to be the right choice.
This is an interesting direction for further exploration \cite{work}.

\vskip7mm
\section*{Acknowledgments}

We would like to thank Calros Cardona, Dimitrios Giataganas,
Wu-Zhong Guo, Yuta Hamada, Hiroyuki Kitamoto, 
Toshifumi Noumi
and Sang-Jin Sin, Gary Shiu
for valuable discussion and  comments.
This work is
supported in part by the National Center for Theoretical Sciences
(NCTS) and the grant 107-2119-M-007-014-MY3 from the Ministry of Science and
Technology of Taiwan.


\appendix

\section{Geodesic equation in AdS space}\label{appgeodesic}

Geodesic equations for a point particle 
in the vacuum AdS space \eqref{ads} are
\begin{align}
{\ddot t}-\frac{2}{y}{\dot t}{\dot y}=0,
\quad
{\ddot x}^{i}-\frac{2}{y}{\dot x}^{i}{\dot y}=0,
\quad
{\ddot y}-\frac{1}{y}({\dot t}^{2}+{\dot y}^{2}
-\delta_{ij}{\dot x}^{i}{\dot x}^{j})=0,
\label{geodesic}
\end{align} 
where dot denotes a derivative 
with respect to the proper time
for massive particles
or the affine parameter for massless particles.
%
There is a constraint equation
\be
\frac{L^{2}}{y^{2}}
(-{\dot t}^{2}
+\delta_{ij}{\dot x}^{i}{\dot x}^{j}+{\dot y}^{2})
=
\begin{cases}
    -1 & (\text{massive}) \\
    0 & (\text{massless})
  \end{cases}
  .
  \label{geoconst}
\ee
Conserved quantities $P_{a}$ 
in the geodesic motion are given by
\be
P_{0}/m 
= -{\bar g}_{00}u^{0}
=
\frac{L^{2}}{y^{2}}{\dot t},
\qquad
P_{i}/m 
= 
{\bar g}_{ij}u^{j}
=
\frac{L^{2}}{y^{2}}\delta_{ij}{\dot x}^{j}.
\label{geocon}
\ee
The energy of particle measured by 
a timelike observer 
whose 4-velocity is $t^{\mu}$ 
is
\be
E=-P_{\mu}t^{\mu}.
\ee
For a timelike Killing vector $t^{\mu}$ in AdS space, 
$t^{\mu}=(-1,0,0,0)$, $E=P_{0}$.
Using \eqref{geoconst} and \eqref{geocon}, 
\eqref{geodesic} becomes
\begin{align}
{\ddot t}
=\frac{2y}{L^{2}}P_{0}{\dot y},
\qquad
{\ddot x}^{i}
=\frac{2y}{L^{2}}P^{i}{\dot y},
\qquad
{\ddot y}
=
\begin{cases}
  \frac{2y^{3}}{L^{4}}
  \Big(\frac{P_{0}^{2}}{m^{2}}
  -\frac{P_{i}^{2}}{m^{2}}
  -\frac{L^{2}}{2y^{2}}\Big) & \ (\text{massive}) \\
  \frac{2y^{3}}{L^{4}}
  (P_{0}^{2}
  -P_{i}^{2}) & \ (\text{massless})
  \end{cases}
  ,
\label{geodesic2}
\end{align} 
where 
$P^{i}=\delta^{ij}P_{j}$, 
$P_{i}^{2}=\delta^{ij}P_{i}P_{j}$. 

For a massive particle, 
from \eqref{geocon} and \eqref{geoconst}, 
we have
\begin{align}
{\dot x}^{i}
&=\frac{y^{2}}{L^{2}}\frac{P_{0}}{m}\frac{dx^{i}}{dt}
=\frac{y^{2}}{L^{2}}\frac{P^{i}}{m}
\notag\\
{\dot y}^{2}
&=
\frac{y^{4}}{L^{4}}\frac{P_{0}^{2}}{m^{2}}
\left(\frac{dy}{dt}\right)^{2}
=
\frac{y^{4}}{L^{4}}
\left(\frac{P_{0}^{2}-P_{i}^{2}}{m^{2}}\right)
-\frac{y^{2}}{L^{2}}
.
\end{align}
It follows that
\begin{align}
x^{i}=\frac{P^{i}}{P_{0}}t +C^{i},
\qquad
y
=\sqrt{\frac{m^{2}L^{2}
+P_{0}^{-2}(P_{0}^{2}-P_{i}^{2})^{2}(t- C_{3})^{2}}
{P^{2}_{0}-P^{2}_{i}}},
\label{massivegeo}
\end{align}
where $C_{i}$ and $C_{3}$ are constants and 
$C_{3}$ determines the time at which $y$ 
takes minimum value. 
Taking $m\to 0$ in \eqref{massivegeo} gives
a solution to the geodesic equation of a massless particle.
The energy of the massless particle is written as
$E=L^{2}y^{-2}{\dot t}$.

\section{Linearized Riemann tensor}
\label{appcurvature}

The Christoffel symbols for the background AdS metric \eqref{ads} is
\be
{\bar \Gamma}^{\rho}_{\mu\nu}
=-\frac{1}{y}
(\delta^{y}_{\mu}\delta^{\rho}_{\nu}+\delta^{y}_{\nu}\delta^{\rho}_{\mu}
-\delta^{\rho}_{y}\eta_{\mu\nu}).
\label{bchris}
\ee
The linearized Riemann tensor is given by 
\begin{align}
R^{\mu(1)}_{\ \alpha\beta\gamma}
&=
\frac12({\bar \nabla}_{\beta}{\bar \nabla}_{\alpha}\gamma^{\mu}_{\ \gamma}
-{\bar \nabla}_{\gamma}{\bar \nabla}_{\alpha}\gamma^{\mu}_{\ \beta}
+{\bar \nabla}_{\gamma}{\bar \nabla}^{\mu}\gamma_{\alpha \beta}
-{\bar \nabla}_{\beta}{\bar \nabla}^{\mu}\gamma_{\alpha \gamma}
+{\bar R}^{\mu}_{\ \rho\beta\gamma}\gamma^{\rho}_{\ \alpha}
-{\bar R}^{\rho}_{\ \alpha\beta\gamma}\gamma^{\mu}_{\ \rho}
).
\label{lriemann}
\end{align}
Substituting \eqref{ads} and \eqref{bchris}
into \eqref{lriemann}, we get
\begin{align}
R^{\mu(1)}_{\ \alpha\beta\gamma}
&=
A^{\mu}_{\ \alpha\beta\gamma}-A^{\mu}_{\ \alpha\gamma\beta}
-{\bar g}^{\mu\rho}{\bar g}_{\alpha\sigma}
(A^{\sigma}_{\ \rho\beta\gamma}-A^{\sigma}_{\ \rho\gamma\beta})
\notag
\\
&\quad
-\frac{1}{2y^{2}}
(
\delta^{\mu}_{\beta}\eta_{\rho\gamma}\gamma^{\rho}_{\ \alpha}
-\delta^{\mu}_{\gamma}\eta_{\rho\beta}\gamma^{\rho}_{\ \alpha}
-\delta^{\rho}_{\beta}\eta_{\alpha\gamma}\gamma^{\mu}_{\ \rho}
+\delta^{\rho}_{\gamma}\eta_{\alpha\beta}\gamma^{\mu}_{\ \rho}
),
\end{align}
where
\begin{align}
A^{\mu}_{\ \alpha\beta\gamma}
&=
\frac12
\left[\partial_{\beta}\partial_{\alpha}\gamma^{\mu}_{\ \gamma}
-\frac{1}{y}
\Big(
\delta^{\mu}_{\alpha}\partial_{\beta}\gamma^{y}_{\ \gamma}
-\frac{y^{2}}{L^{2}}\delta^{\mu}_{ y}\partial_{\beta}\gamma_{\alpha\gamma}
-\delta^{y}_{\gamma}\partial_{\beta}\gamma^{\mu}_{\ \alpha}
+\eta_{\gamma\alpha}\partial_{\beta}\gamma^{\mu}_{\ y}
-\delta^{y}_{\alpha}\partial_{\beta}\gamma^{\mu}_{\ \gamma}
\right.
\notag\\
&\quad
+\eta_{\beta\alpha}\partial_{y}\gamma^{\mu}_{\ \gamma}
+\delta^{\mu}_{\beta}\partial_{\alpha}\gamma^{y}_{\ \gamma}
-\delta^{\mu}_{y}\eta_{\beta\rho}\partial_{\alpha}\gamma^{\rho}_{\ \gamma}
\Big)
+\frac{1}{y^{2}}
\Big(
\delta^{y}_{\beta}\delta^{\mu}_{\alpha}\gamma^{y}_{\ \gamma}
+\delta^{y}_{\alpha}\delta^{y}_{\gamma}\gamma^{\mu}_{\ \beta}
-\frac{y^{2}}{L^{2}}\delta^{\mu}_{\beta}\gamma_{\alpha \gamma}
\notag\\
&\quad
\left.
-\delta^{\mu}_{\beta}\delta^{y}_{\gamma}\gamma^{y}_{\ \alpha}
+\delta^{\mu}_{\beta}\eta_{\alpha\gamma}\gamma^{y}_{\ y}
+\frac{y^{2}}{L^{2}}\delta^{\mu}_{y}\delta^{y}_{\beta}\gamma_{\alpha \gamma}
-\delta^{y}_{\gamma}\eta_{\beta\alpha}\gamma^{\mu}_{\ y}
\Big)
\right]
.
\end{align}

\section{Geodesic and parallel transport equations in perturbed AdS space}
\label{apppert}

Here we consider the geodesic and geodesic deviation equations
 in AdS space with perturbations.
Our analysis is restricted to the first order of the  metric perturbations.

\parabf{Perturbed geodesic and parallel transport equations}

A formalism to construct and solve 
the perturbed geodesic equation in curved spaces 
is given in \cite{Pyne:1993np,Pyne:1995bs}. 
One first decompose the geodesic into 
the background trajectory and perturbation about it
\be
x^{\mu}(\tau)
={\bar x}^{\mu}(\tau)+\delta x^{\mu}(\tau),
\label{xdec}
\ee 
where ${\bar x}^{\mu}$ solves 
the background geodesic equation,
\be
\frac{d^{2}{ {\bar x}}^{\mu}}{d\tau^{2}}
+{\bar \Gamma}^{\mu}_{\alpha\beta}
\frac{d{\bar x}^{\alpha}}{d\tau}
\frac{d{\bar x}^{\beta}}{d\tau}=0,
\ee
with ${\bar \Gamma}^{\mu}_{\alpha\beta}$ 
the Christoffel symbols 
consisting of the background metric,
and $\delta x^{\mu}$ is 
${\cal O}(\gamma_{\mu\nu})$ quantity.
\footnote{
For notation simplicity, 
in the main body 
(section \ref{sectionmemory}) 
we are using $x^{\mu}(\tau)$ 
as the background geodesic 
${\bar x}^{\mu}(\tau)$
as long as there is no confusion.}
Substituting \eqref{xdec} into 
the geodesic equation,
\be
\frac{d^{2}{ { x}}^{\mu}}{d\tau^{2}}
+{\Gamma}^{\mu}_{\alpha\beta}(x)
\frac{d{x}^{\alpha}}{d\tau}
\frac{d{ x}^{\beta}}{d\tau}=0,
\ee
one obtain  
at the first order of perturbation that
\begin{align}
\left(
\delta^{\mu}_{\ \beta}\frac{d^{2}}{d\tau^{2}}
+A^{\mu}_{\ \beta}\frac{d}{d\tau}
+B^{\mu}_{\ \beta}
\right)
{{\delta x}}^{\beta}
=f^{\mu},
\label{pergeodesic}
\end{align}
where
\be
A^{\mu}_{\ \beta}
=2{\bar \Gamma}^{\mu}_{\alpha\beta}
\frac{d{\bar x}^{\alpha}}{d\tau},
\quad
B^{\mu}_{\ \beta}=
\partial_{\beta}({\bar \Gamma}^{\mu}_{\alpha\rho})
\frac{d{\bar x}^{\alpha}}{d\tau}
\frac{d{\bar x}^{\rho}}{d\tau},
\quad
f^{\mu}=
-
{\delta \Gamma}^{\mu}_{\alpha\beta}
\frac{d{\bar x}^{\alpha}}{d\tau}
\frac{d{\bar x}^{\beta}}{d\tau}.
\label{abf}
\ee
In our case of the Poincare coordinates in AdS space, 
$f^{\mu}$ is given by
\be
f^{\mu}
=
-\left(
\partial_{\alpha}h^{\mu}_{\beta}
-\frac12\partial^{\mu}h_{\alpha\beta}
+\frac1y \delta^{\mu}_{y}h_{\alpha\beta}
-\frac1y \eta_{\alpha\beta} h^{y\mu}
\right)
\frac{d{\bar x}^{\alpha}}{d\tau}
\frac{d{\bar x}^{\beta}}{d\tau}
,
\label{fads}
\ee
where $h_{\mu\nu}:= y^{2}L^{-2}\gamma_{\mu\nu}$.

Corresponding to \eqref{xdec},
we also decompose the tetrads into 
its background and perturbation pieces
\be
(e_{\alpha})^{\mu}(\tau)= 
\frac{\partial{x}^{\mu}}{\partial x_{F}^{\alpha}}
\Big|_{\gamma(\tau)}
=
{\bar e}_{\alpha}^{\mu}(\tau)
+{\delta e}_{\alpha}^{\mu}(\tau),
\label{edec}
\ee 
where ${\bar e}_{\alpha}^{\mu}$ solves
the background constraint equation,
${\bar g}_{\mu\nu}({\bar x})
{\bar e}_{\alpha}^{\mu}{\bar e}_{\beta}^{\nu}
=\eta_{\alpha\beta}$.
Note that ${\bar e}_{0}^{\mu}={d{\bar x}^{\mu}}/{d\tau}$
and ${\delta e}_{0}^{\mu}=d{\delta x}^{\mu}/{d\tau}$.
Substituting \eqref{xdec} and \eqref{edec} into 
the constraint equation,
$g_{\mu\nu}(x)(e_{\alpha})^{\mu}(e_{\beta})^{\nu}
=\eta_{\alpha\beta}$,
and the parallel transport equation,
${d}(e_{\alpha})^{\mu}/{d\tau}
+\Gamma^{\mu}_{\ \rho\sigma}(x)
(e_{0})^{\rho}(e_{\alpha})^{\sigma}
=0$, 
we obtain
\bea
&
2{\bar g}_{\mu\nu}
{{\bar e}}^{\mu}_{\alpha}{{\delta e}}^{\nu}_{\beta}
+(\gamma_{\mu\nu}+\partial_{\rho}({\bar g}_{\mu\nu})\delta x^{\rho}
){{\bar e}}^{\mu}_{\alpha}{{\bar e}}^{\nu}_{\beta}
=0,
\label{perconstraint}
\\
&
\frac{d}{d\tau}\delta e_{\alpha}^{\mu}
+ 
{\bar \Gamma}^{\mu}_{\ \rho\sigma}
(\delta e_{0}^{\rho}{\bar e}_{\alpha}^{\sigma}+
{\bar e}_{0}^{\rho}\delta e_{\alpha}^{\sigma})
({\delta \Gamma}^{\mu}_{\alpha\beta}+
\partial_{\beta}({\bar \Gamma}^{\mu}_{\rho\sigma})\delta x^{\beta})
{{\bar e}}^{\rho}_{0}
{{\bar e}}^{\sigma}_{\alpha}
=0.
\label{perparallel}
\eea

\parabf{Solution to the perturbed geodesic equation}

A general solution to \eqref{pergeodesic} 
can be represented by using 
the parallel propagator and 
the Jacobi propagator \cite{Pyne:1995bs}
as follows 
\begin{align}
&
\left(
\begin{array}{c}
  \frac{1}{L}P(\tau_{i},\tau)\delta x(\tau) \\
  \frac{d}{d\tau}(P(\tau_{i},\tau)\delta x(\tau)) 
 \end{array}
 \right)
 \notag\\
&\qquad
=
U(\tau,\tau_{i})
\left(
\begin{array}{c}
 \frac1L \delta x(\tau_{i}) \\
\frac{d}{d\tau'}(P(\tau_{i},\tau')
\delta x(\tau'))|_{\tau'=\tau_{i}} 
 \end{array}
  \right)
  +
  \int^{\tau}_{\tau_{i}}
  d{\bar\tau} U(\tau,{\bar \tau})
  \left(
  \begin{array}{c}
  0_{4}  \\
  P(\tau_{i},{\bar \tau})f({\bar \tau}) 
   \end{array}
\right),
\label{pgeodesicsol}
\end{align}
where we have adopted matrix notation, 
e.g. $P\delta x = P^{\mu}_{\ \alpha} \delta x^{\alpha}$,
$Pf = P^{\mu}_{\ \alpha} f^{\alpha}$,
and $L$ is some length scale which is identified with
the AdS radius in our case. 
The parallel propagator $P(\tau_{1},\tau_{2})$ is 
given by a $4\times 4$ matrix,
\be
P(\tau_{1},\tau_{2})^{\mu}_{\ \nu}
={\cal P}
\exp\Big(-\frac12
\int^{\tau_{1}}_{\tau_{2}} d\tau A(\tau)\Big)^{\mu}_{\ \nu} \ ,
\qquad 
P(\tau_{1},\tau_{2})=P^{-1}(\tau_{2},\tau_{1}),
\label{prapro}
\ee 
where ${\cal P}$ denotes the path ordering and $A$ is given by
$A^{\mu}_{\ \beta}$ in \eqref{abf},
and the Jacobi propagator $U(\tau_{1},\tau_{2})$ is given by
a $8\times 8$ matrix
\be
U(\tau_{1},\tau_{2})
={\cal P}
\exp\left[
 \frac1L
\int^{\tau_{1}}_{\tau_{2}} d\tau
 \left(
  \begin{array}{cc}
  0 & 1_{4} \\
  -P(\tau_{1},{\tau}){\cal R}(\tau)P(\tau,{\tau_{1}}) & 0
   \end{array}
\right)\right],
\label{jacpro}
\ee
where 
${\cal R}^{\mu}_{\ \nu} := {\bar R}^{\mu}_{\ \alpha\nu\beta}
{\bar e}^{\alpha}_{0}{\bar e}_{0}^{\beta}$.
Note that the first term on 
the right hand side of \eqref{pgeodesicsol}
vanishes for a set of initial conditions 
$\delta x^{\mu}(\tau_{i})=d\delta x^{\mu}/d\tau(\tau_{i})=0$.

For the central geodesic $\gamma(\tau)$ in the AdS space,
the parallel and Jacobi propagators are obtained as
\be
P(\tau_{1},\tau_{2})
=
\left(
  \begin{array}{cccc}
  \frac{{\bar y}(\tau_{1})^{2}}{L^{2}}
  \left(1-\frac{{\bar t}(\tau_{1}){\bar t}(\tau_{2})}
  {{\bar y}(\tau_{1}){\bar y}(\tau_{2})}\right)   
   & 0 & 0 & 
 \frac{{\bar y}(\tau_{1})^{2}}{L^{2}}
 \left(\frac{{\bar t}(\tau_{1})}{{\bar y}(\tau_{1})}
 -\frac{{\bar t}(\tau_{2})}{{\bar y}(\tau_{2})}\right) \\
  0 & \frac{{\bar y}(\tau_{1})}{{\bar y}(\tau_{2})} & 0 & 0 \\
  0 & 0 & \frac{{\bar y}(\tau_{1})}{{\bar y}(\tau_{2})} & 0 \\
  \frac{{\bar y}(\tau_{1})^{2}}{L^{2}}
 \left(\frac{{\bar t}(\tau_{1})}{{\bar y}(\tau_{1})}
 -\frac{{\bar t}(\tau_{2})}{{\bar y}(\tau_{2})}\right)
   & 0 & 0 &
    \frac{{\bar y}(\tau_{1})^{2}}{L^{2}}
  \left(1-\frac{{\bar t}(\tau_{1}){\bar t}(\tau_{2})}
  {{\bar y}(\tau_{1}){\bar y}(\tau_{2})}\right)
   \end{array}
\right),
\ee
\be
U(\tau_{1},\tau_{2})
=
\left(
  \begin{array}{cc}
  J\cos\left(\frac{\tau_{1}-\tau_{2}}{L}\right)
   & J\sin\left(\frac{\tau_{1}-\tau_{2}}{L}\right) \\
  -J\sin\left(\frac{\tau_{1}-\tau_{2}}{L}\right) 
  & J\cos\left(\frac{\tau_{1}-\tau_{2}}{L}\right)
   \end{array}
\right)
+
\left(
  \begin{array}{cc}
 1_{4}-J & (1_{4}-J)\frac{\tau_{1}-\tau_{2}}{L}\\
 0 & 1_{4}-J
   \end{array}
\right),
\label{PUads}
\ee
where
\be
J
:=
L^{2}
P(\tau_{1},{\tau}){\cal R}(\tau)P(\tau,{\tau_{1}})
=
\left(
  \begin{array}{cccc}
  -\frac{{\bar y}(\tau_{1})^{2}}{L^{2}}   
   & 0 & 0 & 
 \frac{{\bar y}(\tau_{1}){\bar t}(\tau_{1})}{L^{2}} \\
  0 & 1 & 0 & 0 \\
  0 & 0 & 1 & 0 \\
 -\frac{{\bar y}(\tau_{1}){\bar t}(\tau_{1})}{L^{2}}
   & 0 & 0 &
    \frac{{\bar y}(\tau_{1})^{2}}{L^{2}}
    \end{array}
\right).
\label{Jads}
\ee



\end{document}